\documentclass[aps, prd, twocolumn, 
lengthcheck, superscriptaddress, 
nofootinbib]{revtex4-1}

\usepackage{ulem}
\usepackage{color}
\usepackage{amsfonts}
\usepackage{amsmath}
\usepackage{amssymb}
\usepackage{hyperref}
\usepackage{chngcntr}

\definecolor{ar}{rgb}{1.0, 0.01, 0.24}
\definecolor{al}{rgb}{0.82, 0.1, 0.26}
\definecolor{ev}{rgb}{0.56, 0.0, 1.0}
\def\be{\begin{eqnarray}}\def\ee{\end{eqnarray}}

\usepackage{amsmath,amssymb,amsfonts}
\usepackage{graphicx}
\usepackage{bm}         
\usepackage{hyperref}   
\usepackage{physics}    

\begin{document}




\title{Thermodynamic signatures do not uniquely identify deconfinement in neutron stars}



\author{Yong-Liang Ma}
\email{ylma@nju.edu.cn}
\affiliation{School of Physics, Nanjing University, Nanjing 210093, China}
\affiliation{School of Frontier Sciences, Nanjing University, Suzhou 215163, China}

\author{Jia-Ying Xiong}
\email{xiongjiaying21@mails.ucas.ac.cn}
\affiliation{School of Physics, Nanjing University, Nanjing 210093, China}
\affiliation{School of Frontier Sciences, Nanjing University, Suzhou 215163, China}
\affiliation{School of Fundamental Physics and Mathematical Sciences, Hangzhou Institute for Advanced Study, UCAS, Hangzhou 310024, China}

\date{\today}


\allowdisplaybreaks


\begin{abstract}

The smallness of polytropic index $\Gamma_\varepsilon \lesssim 1.75$ and near conformality of sound velocity $c_s^2 \approx 1/3$ in neutron star matter are usually referred to as signals of the emergence of quark matter. 
We construct a density-resolved nucleonic reference domain in the \((P/P_{\rm free},\Gamma_\varepsilon)\) plane using nucleonic EoSs jointly constrained by nuclear matter properties and neutron star mass, radius and tidal data, and found that the domain extends unambiguously below \(\Gamma_\varepsilon=1.75\). 
Crucially, the nucleonic domain becomes stable against the truncation order only after \(Z_0\) and \(Z_{sym}\) are included, showing that the higher-order density dependence controls the extrapolation from finite nuclei to neutron-star matter. These findings therefore provide concrete targets for finite-nucleus and heavy-ion experiments, while linking terrestrial nuclear physics directly to multimessenger observations.
We compare the domain with smooth equations of state generated by neural networks without phase labels and represented by symbolic regression. Among these reconstructions, \(35.29\%\) have complete trajectories inside the nucleonic domain over \(0.5\leq n/n_0\leq8\). 
Within the EoS sample obtained after minimizing the multimessenger loss, the smallest value of \(\max_n c_s^2(n)\) is 0.38, above the conformal value $1/3$. 
In conclusion, we convert microscopic interpretation into a falsifiable, density-resolved null-hypothesis test and show quantitatively that current observations do not reject the nucleonic null over much of the admissible space of equation of state.

\end{abstract}
\maketitle

\begin{figure*}
    \centering
    \includegraphics[width=0.8\linewidth]{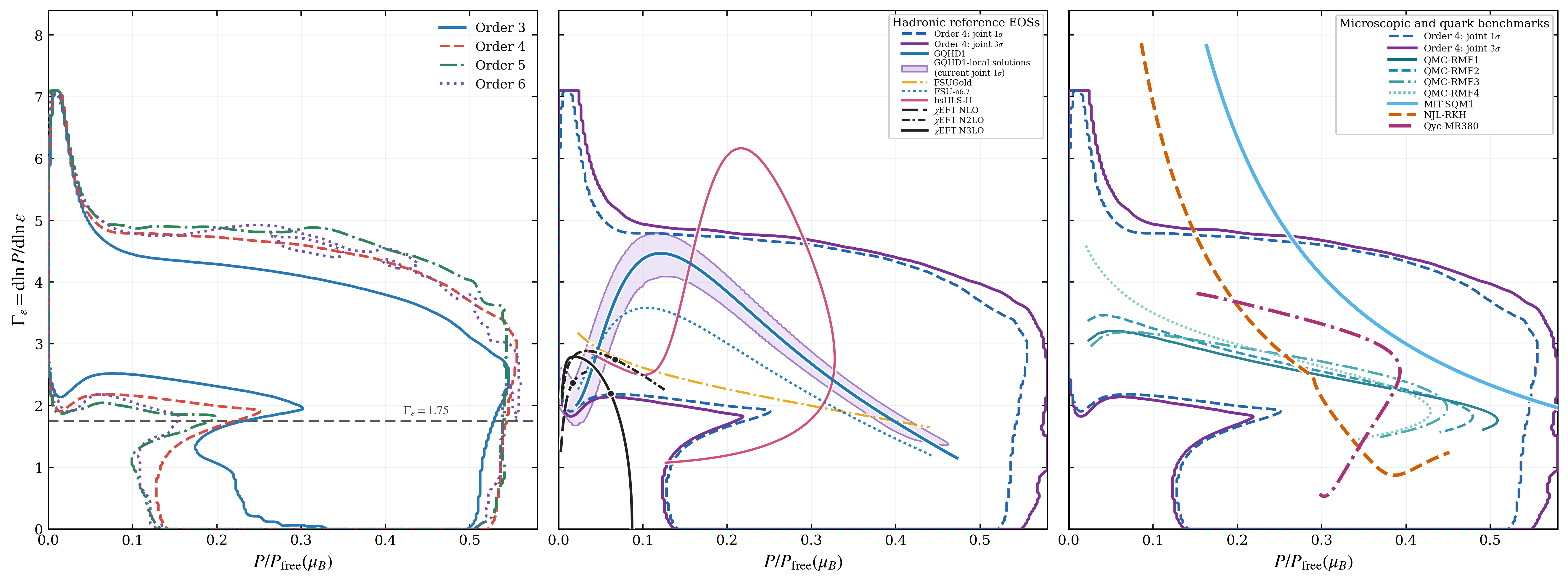}
    \caption{\textbf{Multimessenger posterior structure and Taylor-order convergence in the $(P/P_{\rm free},\Gamma_\epsilon)$ plane.
    The left panel compares the corresponding trajectory-density contours obtained at Taylor orders three to six using the same physical and astronomical conditioning. 
    The right panel illustrates the data-driven thermodynamic regions from the fourth-order \(1\sigma\) and \(3\sigma\) contours in comparison with some representative EOSs.
    The light-purple shaded region represents GQHD solutions satisfying the joint \(3\sigma\) multimessenger criterion. 
    } }
    \label{fig:gp}
\end{figure*}

The nature of nuclear matter (NM) remains a mysterious for several decades.
Neutron-star masses, radii, and tidal deformabilities constrain the equation of state (EoS) of cold dense matter above the saturation density $n_0\approx 0.16$~fm$^{-3}$, but do not directly reveal its microscopic composition~\cite{Antoniadis:2013pzd,LIGOScientific:2020aai,LIGOScientific:2018hze,Antoniadis:2013pzd,Choudhury:2024xbk}. 
Matter containing only nucleons, additional hadronic species or deconfined quarks could produce similar EoSs over the densities reached in stable stars, and hence similar observables~\cite{Baym:2017whm,Chen:2017och,Gil:2018yah,Sorensen:2023zkk}. 
Reconstructing an allowed EoS is therefore not the same as identifying the degrees of freedom that produce it. 
Here we address this distinction by testing whether observationally allowed EOSs remain compatible with the thermodynamic behaviour of nucleonic matter as density increases.


Neutron stars (NSs) connect the physics of finite nuclei and dense strong interaction matter to astrophysical observations~\cite{Horowitz:2014bja}. 
%
The observations of massive NSs with masses around $2M_\odot$~\cite{Demorest:2010bx,Antoniadis:2013pzd} with $M_\odot$ being the solar mass indicate that the EoS of compact star matter requires sufficient pressure at high density. 
However, the measurement of PSR J0030+0451~\cite{Vinciguerra:2023qxq} and PSR J0437-4715~\cite{Choudhury:2024xbk}, and especially PSR J0614-3329~\cite{Mauviard:2025dmd}, soften the EoS at intermediate densities. 
These observations implies that the sound velocity (SV) favors a peak structure at intermediate density~\cite{Altiparmak:2022bke,Marczenko:2022jhl}. 
In addition to the measurement of global properties of NSs, the gravitational wave detection by LIGO gives an independent constraint on the EoS through the tidal deformability~\cite{LIGOScientific:2017vwq,LIGOScientific:2017ync}. 
Taken together, these observations constrain the density dependence of the EoS up to $\sim 10n_0$ with $n_0\approx 0.16$~fm$^{-3}$ being the saturation density but do not uniquely determine the underlying composition~\cite{Mroczek:2023zxo,Chatziioannou:2015uea}. 
Within hadronic matter, it may result from the density evolution of many-body correlations~\cite{Akmal:1998cf} or short-range repulsion~\cite{Gandolfi:2011xu}. 
Similar behaviour is also possible after quark degrees of freedom emerge, either through a smooth crossover~\cite{Masuda:2012ed,Kojo:2014rca} or through a phase transition~\cite{Alford:2013aca,Annala:2019puf} into a sufficiently stiff quark phase. 

Several thermodynamic diagnostics have been proposed as indicators of quark degrees of freedom. One influential prescription associates their onset is the polytropic index, 
\be
\Gamma_\varepsilon=\frac{{\rm d}\ln P}{{\rm d}\ln\varepsilon},
\ee
which should stay below approximately \(1.75\)~\cite{Annala:2019puf,Annala:2023cwx}. Other proposed diagnostics involve the SV, the trace anomaly, and the ratio \(P/P_{\rm free}\) which is sensitive to the number of active degrees of freedom~\cite{Annala:2023cwx,Fujimoto:2022ohj}. These quantities probe the local stiffness and approximate conformality of the EoS rather than its microscopic composition. Purely nucleonic models can display the same features at high density through the density dependence of nuclear interactions, without a phase transition or the appearance of new degrees of freedom~\cite{Takatsy:2023xzf,Malik:2024qjw}. 
A recent generative reconstruction likewise found that the near-conformal thermodynamics together with \(c_s^2>1/3\) (in units of $c=1$ adopted in this paper) in the cores of the heaviest stars are compatible with a gradual hadron--quark crossover but not sufficient to determine the microscopic composition~\cite{Panteleeva:2026zxs}.
Therefore, so far no proposed local thermodynamic diagnostic, alone or jointly, uniquely identifies deconfinement.
In this work, we replace the single-threshold diagnostics with a joint thermodynamic domain constructed from nucleonic EOSs consistent with all the current observations, allowing us to quantify where an inferred EOS departs from the nucleonic reference.

A smooth nucleonic EoS which should be satisfied by matter system without phase transition can be described by the density derivatives around the saturation density \(n_0\) of per particle energy of symmetric NM and symmetry energy. We use these empirical derivatives, rather than a specific hadronic interaction, to construct a broad nucleonic reference ensemble and extend the finite-order expansion to neutron-star densities. Increasing the expansion order then tests whether omitted higher-order terms change the inferred thermodynamic domain. Once this domain becomes stable, it is no longer determined by a particular truncation and can represent, within the adopted nuclear and astronomical constraints, the thermodynamic behaviour of a broad class of smooth nucleonic EoSs.
So we construct a broad EoS ensemble using a Taylor metamodel expanded about the saturation density \(n_0\) and extend it across the densities relevant to NSs. Writing \(x=(n-n_0)/(3n_0)\) and \(\delta=(n_n-n_p)/n\), the energy per-particle is parametrized as
\begin{equation}
E(n,\delta)=\sum_{k=0}^{N}\frac{x^k}{k!}\left(a_k+b_k\delta^2\right),
\end{equation}
where \(a_k=(E_0,0,K_0,Q_0,Z_0,\cdots)\) and \(b_k=(J,L,K_{\rm sym},Q_{\rm sym},Z_{\rm sym},\cdots)\) are empirical parameters describing symmetric NM and the symmetry energy, respectively. Detailed discussions are given in App.~\ref{app:EoS_parametrization_metamodel}. The density dependence of symmetric NM and the symmetry energy is encoded through empirical NM parameters.
The same functional is continued smoothly from saturation to high density, without non-analytic matching, an explicit phase transition or additional hadronic or quark species, thereby defining a fixed-nucleonic null hypothesis.
The core consists of cold, charge-neutral \(npe\) matter in beta equilibrium: particle fractions may evolve with density, but the strong-interaction degrees of freedom remain restricted to nucleons. 
At low densities, the uniform core EoS is matched to the standard BPS crust EoS~\cite{Baym:1971pw} to construct the complete NS EoS.
After imposing thermodynamic stability and causality, we condition the ensemble on three complementary classes of astronomical data listed in Tab.~\ref{tab:observational_data_sources}.

Having established the parametrized nucleonic model and multimessenger constraints, we map the thermodynamic behaviour of nucleonic matter in the \(P/P_{\rm free}\)--\(\Gamma_\varepsilon\) plane, where \(P_{\rm free}(\mu_B)=\mu_B^4/108\pi^2\). 
A large nucleonic reference ensemble, filtered by nuclear and astronomical observation consistency requirements, defines a density-dependent reference region in this plane. 

Because the expansion about \(n_0\) is continued to several times the saturation density, we test the stability of the inferred thermodynamic domain against truncation order. 
The left panel of Fig.~\ref{fig:gp} shows the posterior-weighted two-dimensional trajectory density obtained from third- to sixth-order Taylor expansions of EoSs in the joint \(1\sigma\) multimessenger posterior set over \(1\leq n/n_0\leq8\).
For the third-, fourth-, fifth-, and sixth-order EoS ensembles, respectively, \(100\%\), \(99.99\%\), \(100\%\), and \(100\%\) of the EoSs reach \(\Gamma_\varepsilon<1.75\), demonstrating that this threshold can be attained by purely nucleonic matter and therefore does not by itself imply the emergence of quark degrees of freedom.
In addition, the posterior-weighted trajectory-density outlines from 3-6 orders shown in the left panel of Fig.~\ref{fig:gp} indicating the convergence of the inferred region against truncation order. 
The largest shift occurs between the third- and fourth-order results, whereas the fifth- and sixth-order terms produce only minor corrections.
This shift from third to fourth order indicates that cubic density dependence is insufficient to capture the thermodynamic flexibility of nucleonic matter over NS densities. 
The inclusion of the fourth-order isoscalar and isovector terms, \(Z_0\) and \(Z_{\rm sym}\), is therefore important for stabilizing the extrapolation of smooth nucleonic EoSs to supranuclear densities.
We therefore adopt the fourth-order truncation for the subsequent analysis, as it captures the converged thermodynamic domain without introducing weakly constrained higher-order parameters.
The role of \(Z_0\) and \(Z_{\rm sym}\) in achieving convergence also points to specific opportunities for terrestrial experiments. These coefficients control how the isoscalar and isovector responses evolve as the density moves away from \(n_0\). 
Nuclear compression modes probe the isoscalar density dependence, while neutron skins and dipole polarizabilities along neutron-rich isotopic chains probe its isovector component. Heavy-ion collisions extend this sensitivity towards higher densities. 
Although no single observable determines \(Z_0\) or \(Z_{\rm sym}\) independently, their combined analysis can constrain the relevant parameter combinations and reduce the uncertainty in extrapolating nuclear interactions from finite nuclei to neutron-star densities.

The middle and right panels of Fig.~\ref{fig:gp} compare the fourth-order \(1\sigma\) and \(3\sigma\) trajectory-density regions with representative microscopic and phenomenological EoS predictions, including GQHD1~\cite{Ma:2026kun}, FSUGold~\cite{Piekarewicz:2006bw},TM1~\cite{Sugahara:1993wz},  NL3~\cite{Lalazissis:1996rd}, FSU-\(\delta 6.7\)~\cite{Li:2022okx}, bsHLS-H~\cite{Zhang:2024iye}, chiral-EFT-informed \(QMC-RMF1-4\) family~\cite{Alford:2022bpp}, the microscopic \(NLO,\ N^2LO,\ N^3LO\) chiral-EFT calculation~\cite{Gottling:2025ohe}, MIT-SQM1~\cite{Lattimer:2000nx}, NJL-RKH~\cite{Rehberg:1995kh}, and quarkyonic MR380~\cite{McLerran:2018hbz}.
Together, these models cover a broad range of nuclear interactions and high-density descriptions, including phenomenological nuclear models, microscopic many-body calculations, quark-matter EoSs, and an EoS containing both nucleonic and quark degrees of freedom.
Each has been calibrated to finite-nuclei or NM properties and applied to NSs.
The nucleonic GQHD1, FSUGold, FSU-$\delta6.7$, and QMC--RMF1--4 trajectories lie predominantly within the thermodynamic domain demonstrate that the inferred thermodynamic behaviour is not an artifact of a particular parametrization, but is shared by a broad class of smooth nucleonic EoSs.
The light-purple region, by contrast, consists of GQHD solutions satisfying the same NM requirements and joint \(3\sigma\) multimessenger criterion, and therefore provides an explicit realization within an independently parametrized hadronic Lagrangian framework.
Crucially, this GQHD region extends below \(\Gamma_\epsilon=1.75\). 
The more restricted extent of the GQHD region is expected because a specific hadronic Lagrangian explores only a lower-dimensional subset of the broader metamodel space. Accordingly, the GQHD region should not be interpreted as the complete domain of hadronic matter, and regions not covered by it cannot be identified as nonhadronic solely on this basis.
It therefore constitutes a constructive hadronic counterexample to interpreting \(\Gamma_\epsilon<1.75\) as a unique signature of quark degrees of freedom, rather than merely providing a consistency check of the metamodel inference. 
The partial departure of bsHLS-H is consistent with the tension of its relatively large $K_0$ and $L$ with the empirical ranges adopted here. The $\mathrm{NLO}$, $\mathrm{N}^{2}\mathrm{LO}$, and $\mathrm{N}^{3}\mathrm{LO}$ chiral-EFT results provide an independent low-density anchor. The terminal decrease of the $\mathrm{N}^{3}\mathrm{LO}$ curve occurs near the upper limit of the EFT calculation, where truncation and derivative uncertainties increase, and should not be interpreted as a phase-transition signal.
Turning to the quark-matter models, the MIT-SQM1 and NJL-RKH trajectories partly overlap with the nucleonic reference domain and partly extend beyond it. 
The hadronic and quark predictions are therefore not separated by a sharp boundary in the \(\left(P/P_{\rm free},\Gamma_\epsilon\right)\) plane. 
This overlap leads us to conjecture that it may correspond to a region in which nucleonic and quark degrees of freedom coexist.
To examine this interpretation, we further include Qyc-MR380, which contains both degrees of freedom by construction. Its trajectory lies within the proposed coexistence region, providing an explicit realization of this possibility.

After conditioning the fourth-order EoS ensemble on the combined mass, radius, and tidal-deformability likelihoods, we obtain the posterior constraints on the NM parameters summarized in Tab.~\ref{tab:nuclear_parameters}. These intervals are conditional on the Taylor metamodel and its adopted parameter priors. Propagating the selected parameter ensemble through the EoS, Tolman–Oppenheimer–Volkoff (TOV), and tidal calculations yields the corresponding matter and NS properties shown in Tab.~\ref{tab:NS_properties}.

To explore high-density thermodynamic behaviour without assigning a microscopic phase, we combine neural network (NN) reconstruction with symbolic regression (SR).
NNs provide the flexibility needed to generate a broad ensemble of smooth barotropic relations \(P(\varepsilon)\) satisfying thermodynamic stability, causality and the adopted astrophysical constraints. We optimize the parameters of each NN EoS by minimizing Loss which is defined in Eq.~\eqref{eq:Losstotal}.
SR searches for compact analytic representations of the reconstructed curves and retains only those satisfying the prescribed accuracy and complexity criteria.
Each accepted EoS satisfies thermodynamic stability, causality and the adopted astronomical selection, but carries no microscopic composition label. The resulting solutions may therefore admit hadronic, quark-dominated or crossover interpretations.
Conclusions drawn from this ensemble are consequently not tied to a specific microscopic model and are broadly applicable within the class of smooth, single-branch EOSs considered here.
We project the portions realized inside stable compact-star configurations onto the \(\left(\Gamma_\varepsilon,P/P_{\rm free}\right)\) plane and determine the fraction lying within the nucleonic reference region. 
Starting from an initial ensemble of \(379{,}170\) NN EoS candidates, we minimize the combined physical and astronomical loss and retain \(103{,}555\) models which satisfy the approximate joint-\(1\sigma\) criterion, \(T_{\beta{\rm eq}}\leq\chi_8^2(0.6827)\), and for which SR yields a valid smooth representation.
This common NN--SR ensemble is used throughout the subsequent analysis. Some details of SR related to the present work are discussed in App.~\ref{app:multimessenger_likelihood} and App.~\ref{app:EoS_parametrization_NN}.

Of the \(103{,}555\) valid symbolic EoS reconstructions, \(92.65\%\) of all sampled trajectory points lie within the reference region. At the level of complete EoSs, \(65.47\%\) have at least \(90\%\) of their trajectory points are covered by the fourth-order nucleonic reference region shown in Fig.~\ref{fig:gp}, while \(45.72\%\) remain entirely within this reference region. 
Thus, much of the thermodynamic behaviour exhibited by the astrophysically admissible smooth EoSs is compatible with a purely nucleonic description.

\begin{figure}
    \centering
    \includegraphics[width=0.9\linewidth]{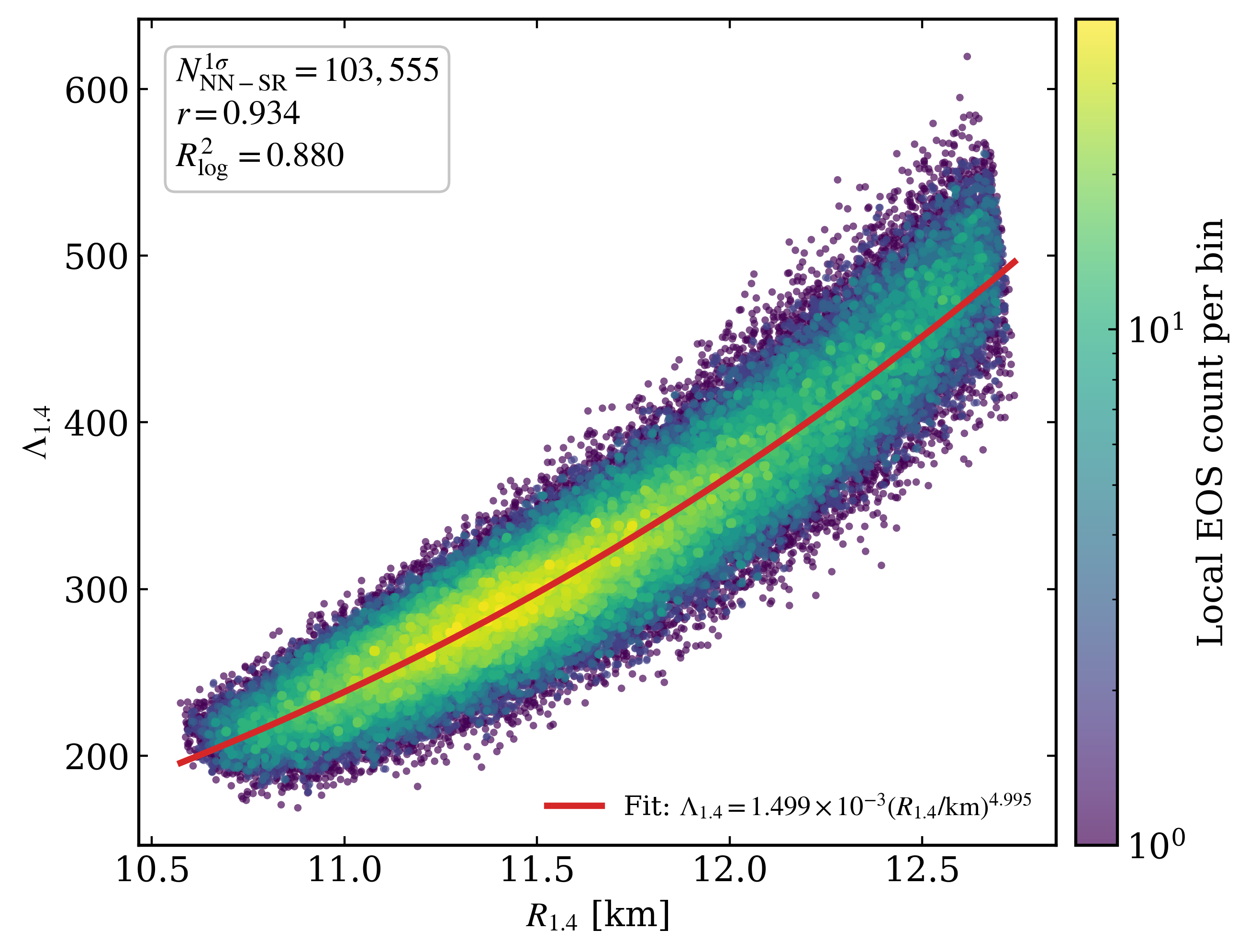}
    \caption{Radius–tidal-deformability correlation in the symbolically represented EoS ensemble}
    \label{fig:RL1.4}
\end{figure}

Fig.~\ref{fig:RL1.4} displays the correlation between the radius \(R_{1.4}\) and the dimensionless tidal deformability \(\Lambda_{1.4}\) of a \(1.4\,M_\odot\) NS for the \(103{,}555\) EoSs in the completed SR ensemble. The color scale represents the local occurrence count on a logarithmic scale, with brighter colors indicating more densely populated regions. A strong positive correlation is observed, with a Pearson coefficient of \(r=0.934\). A power-law relation is obtained by performing a linear regression in logarithmic space,
\begin{equation}
\ln\Lambda_{1.4}=-6.5028+4.9946\ln\left(\frac{R_{1.4}}{\mathrm{km}}\right).
\end{equation}
with a coefficient of determination \(R_{\log}^{2}=0.880\). 
This value indicates that the radius-based power law accounts for \(88.0\%\) of the variance in \(\ln\Lambda_{1.4}\) relative to a constant-mean model. The fitted exponent is close to the fifth-power dependence expected from \(\Lambda_{1.4}=(2/3)k_{2,1.4}C_{1.4}^{-5}\) where \(C_{1.4}=GM_{1.4}/R_{1.4}\) is the compactness of a \(1.4\,M_\odot\) NS, \(k_{2,1.4}\) is its quadrupolar tidal Love number, and \(G\) is the gravitational constant. The remaining scatter reflects EoS-dependent variations in the tidal Love number and the internal density profile.
\begin{figure*}
    \centering
    \includegraphics[width=1\linewidth]{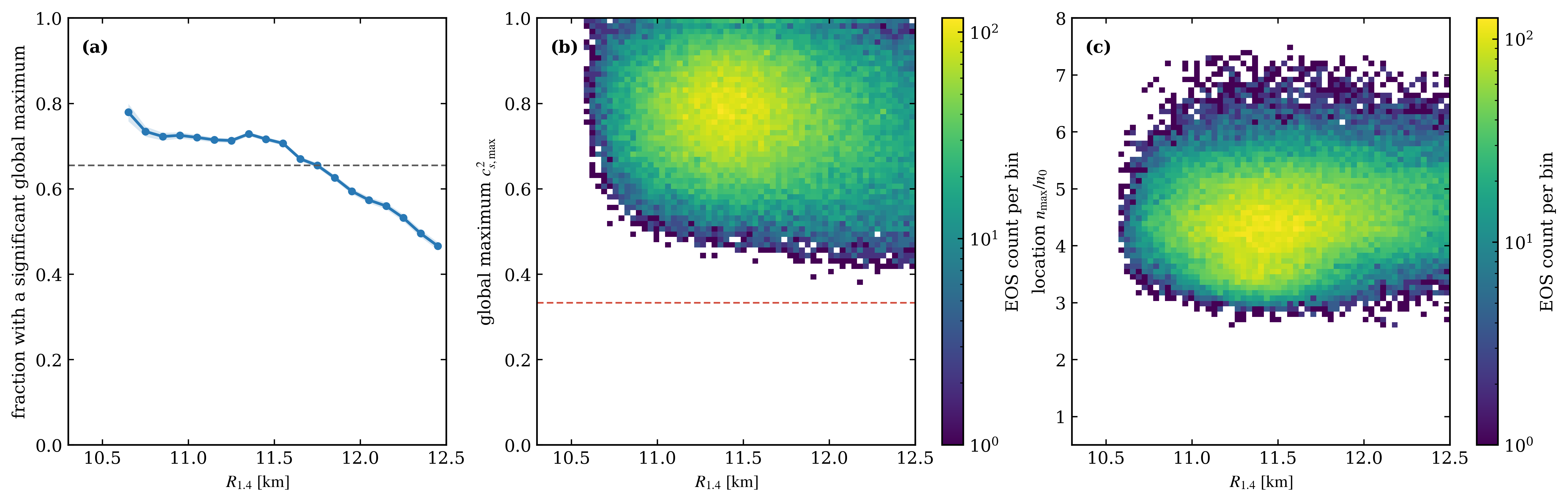}
    \caption{Radius dependence of SV maxima in the multimessenger-constrained EoS ensemble. The left panel shows the fraction of EoSs exhibiting a significant global maximum in the SV as a function of $R_{1.4}$ with the horizontal dashed line denotes the mean peak fraction of the complete EoS ensemble. The middle panel presents the joint distribution of the peak height $c_{s,\max}^2$ and \(R_{1.4}\), with the red dashed line marking the conformal benchmark $c_s^2=1/3$. The right panel shows the joint distribution of the corresponding peak location $n_{\max}/n_0$ and \(R_{1.4}\).}
    \label{fig:PR14}
\end{figure*}

Among the \(103{,}555\) EoSs analyzed, \(67{,}822\) exhibit a significant SV peak in \(c_s^2\), corresponding to \(65.49\%\) of the complete ensemble. 
Here, a peak is classified as significant only if the global maximum occurs inside the sampled density interval and is followed by a decrease in $c_s^2$ of at least $0.05$.
As shown in the left panel of Fig.~\ref{fig:PR14}, the peak fraction remains approximately \(0.71–0.83\) for $10.5\lesssim R_{1.4}\lesssim11.6~\mathrm{km}$. It then decreases systematically with increasing radius, reaching approximately $0.47$ near $R_{1.4}=12.5~\mathrm{km}$. EoSs producing more compact $1.4\,M_{\odot}$ NSs are therefore more likely to exhibit a pronounced nonmonotonic SV structure than those predicting larger radii.
The middle panel shows that, among EoSs with a significant maximum, the peak height is predominantly distributed over $c_{s,\max}^2\simeq0.6$--$0.9$, with the great majority of the maxima lying above the conformal benchmark $c_s^2=1/3$. 
The right panel indicates that the maxima occur primarily at $3\lesssim n_{\max}/n_0\lesssim6$, with the highest occurrence density around $n_{\max}\simeq(3.5\text{--}5)n_0$. These results suggest that compact NS solutions allowed by the current multimessenger constraints frequently require substantial EoS stiffening at intermediate densities, followed by relative softening at higher densities.
It characterizes a nonmonotonic evolution of the macroscopic stiffness of dense matter within the adopted EoS representation and observational constraints.

The left $34.51\% (35,733)$ multimessenger-constrained EoSs do not exhibit a significant SV peak over $0.5\leq n/n_0\leq8$. 
Nevertheless, all EoSs in this subset reach $c_{s,\max}^2>1/3$, with a median and 16th--84th percentile interval of $0.845^{+0.107}_{-0.141}$. These solutions span a broad range of $R_{1.4}$, demonstrating that the absence of a resolved SV peak is not restricted to a narrow canonical-radius interval. This result reveals a weak positive correlation between $M_{\max}$ and $c_{s,\max}^2$, with a Pearson correlation coefficient of $r=0.263$. EoSs supporting larger maximum masses therefore tend statistically to attain larger SV at high density. 

In summary, the above results indicate that, within the phase-transition-free EoS class explored in this work, no EoS satisfying $c_s^2\leq1/3$ throughout $0.5\leq n/n_0\leq8$ can simultaneously support the observed massive NSs and reproduce the inferred radii and tidal deformabilities. 
During NN optimization, EoSs with $c_{s,\max}^2<1/3$ could arise, mainly near the boundary of the approximate joint-$1\sigma$ compatibility region. As the observational loss is minimized, the search moves toward better-fitting solutions with larger maximum sound velocity. For the optimized EoSs retained in this work, we obtain a numerical lower bound of $c_{s,\max}^2\gtrsim0.38$. This value remains appreciably above the conformal SV $c_s^2 \simeq 1/3$.
The multimessenger data therefore favor an exceedance of the conformal SV benchmark, $c_s^2>1/3$, over some interval of supranuclear density. This conclusion is highly general within the class considered here. NNs explore continuous, smooth, causal and thermodynamically stable single-branch barotropes without fixing the microscopic composition, interaction or analytic form of the high-density EoS. 
Each candidate is evaluated against a likelihood combining massive-pulsar masses, NICER mass–radius measurements and gravitational-wave tidal data. 
Within this broad phase-transition-free ensemble, the occurrence of $c_s^2>1/3$ thus emerges statistically from the joint astrophysical constraints rather than from a prescribed microscopic model or analytic ansatz.

\begin{figure*}
    \centering
    \includegraphics[width=1\linewidth]{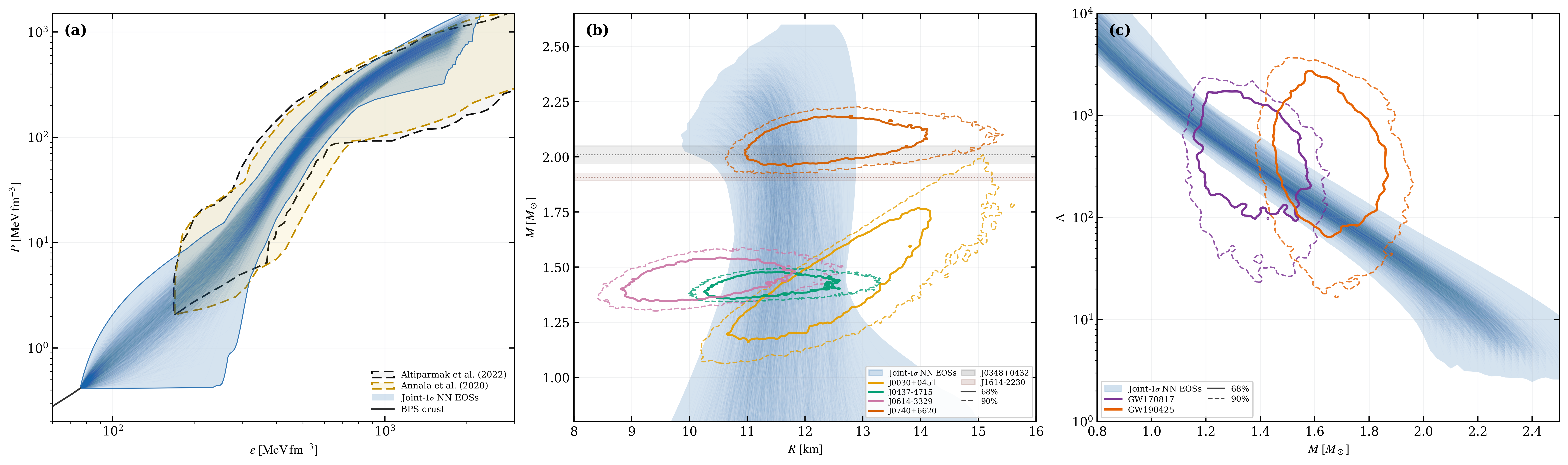}
    \caption{M-R structure and tidal response of the smooth high-density EoS continuations.}
    \label{fig:MR}
\end{figure*}

Fig.~\ref{fig:MR} presents the pressure–energy-density relations of the EoSs selected by the joint \(1\sigma\) multimessenger constraint directly to the complete ensemble of \(379{,}170\) valid symbolic EoSs. 
In the left panel of Fig.~\ref{fig:MR}, the blue region therefore represents the thermodynamic domain accessible to such smooth, phase-transition-free continuations that remain consistent with the adopted multimessenger data. 
For comparison, the regions obtained by Altiparmak {\it et al}.~\cite{Altiparmak:2022bke} and Annala {\it et al}.~\cite{Annala:2019puf} are enclosed by the black and yellow dashed boundaries, respectively. 
Their broad overlap with our region reflects the physical and astrophysical requirements shared by these analyses. 
The remaining differences arise mainly from the high-density parametrization: unlike our globally smooth construction, the piecewise interpolation schemes used in the earlier studies permit rapid changes over narrow density intervals and can accommodate strong crossovers or phase-transition-like behaviour. The agreement therefore holds at the level of the projected \(P(\varepsilon)\) relation and does not imply identical derivative behaviour, microscopic composition or phase structure.

The joint likelihood combines the NICER mass-radius constraints for PSRs J0030$+$0451~\cite{Kini:2026rjx}, J0437$-$4715~\cite{Choudhury:2024xbk}, J0614$-$3329~\cite{Mauviard:2025dmd}, and J0740$+$6620~\cite{Salmi:2024aum}, the massive-pulsar measurements of PSRs J0348$+$0432~\cite{Antoniadis:2013pzd} and J1614$-$2230~\cite{NANOGRAV:2018hou}, and the mass-tidal-deformability information from GW170817~\cite{LIGOScientific:2018hze} and GW190425~\cite{LIGOScientific:2020aai}.
The middle and right panel of Fig.~\ref{fig:MR} show the compact-star sequences that remain compatible with the astronomical data afterthe symbolic reconstructions themselves, rather than only their parent NNs, have been subjected to the joint likelihood.

The complete sample of SR EoSs satisfying the joint-$1\sigma$ criterion spans
\begin{equation}
2.010\leq \frac{M_{\max}}{M_\odot}\leq2.636.
\end{equation}
At $1.4,M_\odot$, the full ranges of the selected EoSs predict
\begin{equation}
10.575\leq \frac{R_{1.4}}{{\rm km}}\leq12.739,\qquad168.8\leq\Lambda_{1.4}\leq619.4.
\end{equation}
While at $2.0,M_\odot$ they become
\begin{equation}10.15\leq \frac{R_{2.0}}{{\rm km}}\leq12.89,\qquad6.970\leq\Lambda_{2.0}\leq67.42.
\end{equation}

\begin{table*}[t]
\centering
\caption{Nuclear-matter parameters of the $70{,}308$ fourth-order Taylor EoSs selected by the approximate joint-$1\sigma$ criterion, $T_{\rm had}\leq\chi_{13}^{2}(0.6827)$, compared with representative hadronic models. The reference-model values are calculated using the same derivative definitions as those employed in the Taylor expansion. All values are quoted to four significant figures.}
\label{tab:nuclear_parameters}
\small
\resizebox{\textwidth}{!}{%
\begin{tabular}{lccccccc}
\hline
Parameter & Posterior median $\pm1\sigma$ & $68\%$ credible interval
& FSUGold~\cite{Todd-Rutel:2005yzo}
& bsHLS-H~\cite{Zhang:2024iye}
& bsHLS-L~\cite{Zhang:2024iye}
& GQHD2~\cite{Ma:2026kun} & Unit \\
\hline
$n_0$ & $0.1609^{+0.004294}_{-0.004167}$ & $[0.1567,\,0.1652]$ & $0.1484$ & $0.1592$ & $0.1592$ & $0.1550$ & $\mathrm{fm}^{-3}$ \\
$E_0$ & $-16.20^{+0.8313}_{-0.8319}$ & $[-17.03,\,-15.37]$ & $-16.30$ & $-16.00$ & $-16.00$ & $-16.02$ & $\mathrm{MeV}$ \\
$K_0$ & $232.0^{+24.97}_{-26.22}$ & $[205.8,\,257.0]$ & $230.0$ & $283.8$ & $232.3$ & $213.9$ & $\mathrm{MeV}$ \\
$Q_0$ & $-72.62^{+177.1}_{-162.0}$ & $[-234.7,\,104.5]$ & $-523.3$ & $-555.4$ & $-779.4$ & $-470.4$ & $\mathrm{MeV}$ \\
$Z_0$ & $-30.56^{+155.7}_{-183.3}$ & $[-213.9,\,125.2]$ & $2823$ & $-504.8$ & $2.067\times10^{5}$ & $4419$ & $\mathrm{MeV}$ \\
$J$ & $30.35^{+1.453}_{-1.427}$ & $[28.92,\,31.80]$ & $32.59$ & $29.23$ & $30.49$ & $33.13$ & $\mathrm{MeV}$ \\
$L$ & $48.55^{+10.93}_{-11.96}$ & $[36.59,\,59.47]$ & $60.52$ & $72.30$ & $86.62$ & $49.76$ & $\mathrm{MeV}$ \\
$K_{\mathrm{sym}}$ & $-11.38^{+72.03}_{-87.45}$ & $[-242.6,\,-44.15]$ & $-51.31$ & $-49.91$ & $-95.12$ & $-278.5$ & $\mathrm{MeV}$ \\
$Q_{\mathrm{sym}}$ & $314.8^{+321.2}_{-383.6}$ & $[164.1,\,702.8]$ & $424.3$ & $35.24$ & $-8704$ & $395.2$ & $\mathrm{MeV}$ \\
$Z_{\mathrm{sym}}$ & $676.6^{+583.7}_{-787.5}$ & $[-89.82,\,1.260\times10^{3}]$ & $-6331$ & $-971.1$ & $-6.994\times10^{5}$ & $3055$ & $\mathrm{MeV}$ \\
\hline
\end{tabular}%
}
\end{table*}

\begin{table}[t]
\centering
\caption{EoS and neutron-star properties obtained from the $70{,}308$ fourth-order Taylor EoSs selected by the approximate joint-$1\sigma$ criterion, $T_{\rm had}\leq\chi_{13}^{2}(0.6827)$. 
Properties of $2.0,M_\odot$ stars are evaluated using the $67{,}501$ EoSs that support a stable configuration at this mass and yield a finite central density. All values are quoted to four significant figures.}
\label{tab:NS_properties}
\begin{tabular}{lccc}
\hline
Quantity & Median $\pm 1\sigma$  & $68\%$ credible interval & Unit \\
\hline
$M_{\max}$ & $2.294$ & $[1.923,\,2.724]$ & $M_\odot$ \\
$R(M_{\max})$ & $11.39$ & $[9.475,\,12.45]$ & $\mathrm{km}$ \\
$R_{1.4}$ & $12.45$ & $[10.48,\,13.12]$ & $\mathrm{km}$ \\
$R_{2.0}$ & $12.32$ & $[9.774,\,13.28]$ & $\mathrm{km}$ \\
$\Lambda_{1.4}$ & $520.3$ & $[156.1,\,749.9]$ & dimensionless \\
$\Lambda_{2.0}$ & $43.82$ & $[4.852,\,83.91]$ & dimensionless \\
$n_c(2M_\odot)$ & $0.5645$ & $[0.4175,\,1.281]$ & $\mathrm{fm}^{-3}$ \\
\hline
\end{tabular}
\end{table}

In conclusion, no presently proposed local thermodynamic diagnostic, alone or jointly, uniquely identifies deconfinement.
We identify a density-resolved thermodynamic domain accessible to smooth nucleonic matter consistent with current nuclear and multimessenger constraints. 
The agreement among hadronic EoSs constructed within different theoretical frameworks shows that this region reflects common hadronic thermodynamic behaviour rather than an artifact of a specific functional form.
Its boundary becomes stable against the truncation order only after the fourth-order coefficients \(Z_0\) and \(Z_{\rm sym}\) are included, showing that higher-order density dependence is essential for connecting finite nuclei to neutron-star matter. 
These coefficients therefore provide concrete targets for finite-nucleus and heavy-ion experiments.

Applying this reference domain to an independently generated ensemble of smooth EoSs shows that a purely nucleonic interpretation remains thermodynamically viable for a substantial fraction of the solutions allowed by current observations. In the SR ensemble, \(45.72\%\) of the EoS trajectories remain entirely within the nucleonic domain over \(1\leq n/n_0\leq8\), while \(65.47\%\) are covered over at least \(90\%\) of this interval. The observed combination of compact radii and large maximum masses can therefore be reproduced by smooth single-branch EoSs without imposing an explicit phase transition. This does not exclude phase transitions but shows that they are not required by the current data within the EoS class considered here.

The same ensemble reveals that compact intermediate-mass stars frequently require pronounced stiffening at supranuclear density. Approximately two-thirds of the accepted EoSs develop a significant maximum in \(c_s^2\), typically between \(3n_0\) and \(6n_0\), with peak values predominantly above the conformal benchmark. More importantly, direct constrained optimization gives
\be
\min_{\mathcal E_{\rm adm}}\!\left[\max_n c_s^2(n)\right]=0.38,
\ee
showing that \(c_s^2>1/3\) is required somewhere within the smooth single-branch EoS class considered.
The exceedance of the conformal benchmark and the appearance of a SV maximum therefore encode the stiffness demanded by the combined observations, but do not by themselves identify a change in microscopic degrees of freedom.
Unlike the \(0.03 \%\) sub-conformal fraction reported in Ref.\cite{Altiparmak:2022bke}, none of our accepted EOSs remains below \(c_s^2=1 / 3\), because our construction imposes global smoothness.

Our results separate two questions that are often conflated. Multimessenger data can test an explicitly defined nucleonic null hypothesis, but a departure from that hypothesis does not determine the alternative phase. The present conclusions remain conditional on the nuclear metamodel, its priors and the restriction to smooth single-branch EoSs; pressure discontinuities and disconnected stable branches associated with a first-order transition are not included. Extending the same construction to independently calibrated quark and hybrid-matter domains using, e.g., NJL~\cite{Hatsuda:1994pi}, PNJL~\cite{Ratti:2006ka}, and quarkyinic~\cite{McLerran:2018hbz} models, together with improved radius and tidal measurements, will be necessary to turn thermodynamic compatibility into a quantitative discrimination of NS core composition. 

\acknowledgments
Y.~L. M. is supported in part by the National Science Foundation of China (NSFC) under Grant No. 12547104, the National Key R\&D Program of China under Grant No. 2021YFC2202900 and Gusu Talent Innovation Program under Grant No. ZXL2024363.

\section*{Data availability}
The data that support the findings of this article, including the nuclear-matter parameter sets, trained neural-network weights, symbolic expressions, and equations of state, are openly available in the GitHub repository at \url{https://github.com/Xiongjiaying-phy/multimessenger-nuclear-eos}.


\begin{widetext}

\appendix 

\counterwithin{table}{section}
\counterwithin{figure}{section}

\section{Multimessenger likelihood and observational conditioning}
\label{app:multimessenger_likelihood}

For every EoS, denoted collectively by the parameters or functional degrees of freedom $\boldsymbol\theta$, we solve the TOV and tidal-perturbation equations and retain the stable compact-star branch. This gives the EoS-dependent functions
\begin{equation}
 R_{\boldsymbol\theta}(M),\qquad
 \Lambda_{\boldsymbol\theta}(M),\qquad
 0<M\leq M_{\max}(\boldsymbol\theta),
\end{equation}
where $R$, $\Lambda$, and $M_{\max}$ are the radius, dimensionless tidal deformability, and maximum mass, respectively. The astronomical data set $d$ combines four pulse-profile mass--radius measurements, two independent massive-pulsar measurements, and the binary-neutron-star events GW170817 and GW190425 summarized in Tab.~\ref{tab:observational_data_sources}. Conditional independence of these data sets is assumed after accounting for measurements already embedded in the released joint products. 
The combined likelihood is then decomposed as
\begin{equation}
\mathcal L_{\rm joint}(\boldsymbol{\theta})=\mathcal L_{\rm NM}\ \mathcal L_{\rm astro}(\boldsymbol\theta).
\label{eq:joint_likelihood}
\end{equation}
The astronomical likelihood is therefore factorized as
\begin{equation}
    \label{eq:astro}
    \begin{aligned}
     \mathcal L_{\rm astro}(\boldsymbol\theta)
     ={}& \prod_{d\in\mathrm{NICER,\ GW}}
           \mathcal L_{d}^{\rm 2D}(\boldsymbol\theta)
           \prod_{a\in\mathrm{massive}}
     \mathcal L_a^{\rm 1D}(\boldsymbol\theta).
    \end{aligned}
\end{equation}
where \(\boldsymbol{\theta}\) denotes the EoS parameters, \(d\) collectively labels the four NICER mass--radius measurements for PSRs J0030$+$0451, J0437$-$4715, J0614$-$3329, and J0740$+$6620, together with the two mass--tidal-deformability measurements from GW170817 and GW190425, \(a\) labels the two independent massive-pulsar mass measurements for PSRs J0348$+$0432 and J1614$-$2230. 
For each GW event, the effective \((M,\ln\Lambda)\) distribution is constructed by transforming the released \((\mathcal M_c,q,\Lambda_1,\Lambda_2)\) samples to the component masses and pooling the two component-star projections.
The quantities \(\mathcal L_d^{\rm 2D}\) and \(\mathcal L_a^{\rm 1D}\) denote the corresponding two- and one-dimensional likelihood contributions, respectively. The public data are summarized in Tab.~\ref{tab:observational_data_sources}.

\begin{table*}[t]
\centering
\caption{Astronomical data sets included in the baseline multimessenger likelihood.}
\label{tab:observational_data_sources}
\small
\begin{tabular}{lccc}
\hline
Name & Observable/constraint type& Reference & Data/source \\
\hline
PSR J0030$+$0451& NICER/XMM joint \(M\)--\(R\) likelihood & ~\cite{Kini:2026rjx} & \href{https://zenodo.org/records/20271881}{Zenodo 20271881} \\
PSR J0437$-$4715 & NICER/XMM+radio joint \(M\)--\(R\) likelihood& ~\cite{Choudhury:2024xbk} & \href{https://zenodo.org/records/13766753}{Zenodo 13766753} \\
PSR J0614$-$3329 & NICER/XMM+radio joint \(M\)--\(R\) likelihood& ~\cite{Mauviard:2025dmd} & \href{https://zenodo.org/records/17380576}{Zenodo 17380576} \\
PSR J0740$+$6620 & NICER/XMM+radio joint \(M\)--\(R\) likelihood& ~\cite{Salmi:2024aum} & \href{https://zenodo.org/records/10519473}{Zenodo 10519473} \\
PSR J0348$+$0432 & NICER/XMM+radio joint \(M\)--\(R\) likelihood& ~\cite{Antoniadis:2013pzd} & \href{https://arxiv.org/abs/1304.6875}{arXiv:1304.6875} \\
PSR J1614$-$2230 & NICER/XMM+radio joint \(M\)--\(R\) likelihood& ~\cite{NANOGrav:2017wvv} & \href{https://arxiv.org/abs/1801.01837}{arXiv:1801.01837} \\
GW170817 & GW mass-tidal likelihood \((\mathcal M_c,q,\Lambda_1,\Lambda_2)\) & ~\cite{LIGOScientific:2018hze} & \href{https://dcc.ligo.org/LIGO-P1800061/public}{LIGO-P1800061-v11} \\
GW190425 & GW mass-tidal likelihood \((\mathcal M_c,q,\Lambda_1,\Lambda_2)\) & ~\cite{LIGOScientific:2020aai} & \href{https://dcc.ligo.org/LIGO-P2000026/public}{LIGO-P2000026-v2} \\
\hline
\end{tabular}
\end{table*}

For each data set \(d\), we introduce two observables \(x_d\) and \(y_d\),
defined by
\begin{equation}
(x_d,y_d)=
\begin{cases}
(M,R),&d\in\mathrm{NICER},\\[1mm]
(M,\ln \Lambda),&d\in\mathrm{GW}.
\end{cases}
\label{eq:2D_observable_definition}
\end{equation}

We approximate each two-dimensional observational constraint \(d\) by a correlated bivariate Gaussian with central values \((\mu_{x,d},\mu_{y,d})\), standard deviations \((\sigma_{x,d},\sigma_{y,d})\), and correlation coefficient \(\rho_d\). The corresponding Gaussian-ellipse statistic is
\begin{equation}
\chi_d^2(x_d,y_d)=\frac{1}{1-\rho_d^2}\left[\left(\frac{x_d-\mu_{x,d}}{\sigma_{x,d}}\right)^2+\left(\frac{y_d-\mu_{y,d}}{\sigma_{y,d}}\right)^2-2\rho_d\left(\frac{x_d-\mu_{x,d}}{\sigma_{x,d}}\right)\left(\frac{y_d-\mu_{y,d}}{\sigma_{y,d}}\right)\right],
\label{eq:unified_2D_chi2}
\end{equation}
with the fitted Gaussian parameters listed in Table~\ref{tab:2d_gaussian_parameters}. Then, one can express the corresponding peak-normalized likelihood as
\begin{equation}
\mathcal L_{d,\mathrm{obs}}(x_d,y_d)=\exp\left[-\frac{1}{2}\chi_d^2(x_d,y_d)\right],\qquad \mathcal L_{d,\mathrm{obs}}(\mu_{x,d},\mu_{y,d})=1.
\label{eq:unified_2D_likelihood}
\end{equation}
The likelihood contribution from data set \(d\) and its associated statistic are
\begin{equation}
\mathcal L_d^{\rm 2D}(\boldsymbol{\theta})=\operatorname*{max}_{M\in\mathcal S_{\boldsymbol{\theta}}}\mathcal L_{d,\mathrm{obs}}\!\left[M,y_{\boldsymbol{\theta},d}(M)\right],\qquad T_d^{\rm 2D}(\boldsymbol{\theta})=-2\ln\mathcal L_d^{\rm 2D}(\boldsymbol{\theta})=\operatorname*{min}_{M\in\mathcal S_{\boldsymbol{\theta}}}\chi_d^2\!\left[M,y_{\boldsymbol{\theta},d}(M)\right].
\label{eq:unified_2D_profile}
\end{equation}
where
\begin{equation}
\mathcal S_{\boldsymbol{\theta},d}^{\rm 2D}\equiv
\begin{cases}
\left\{\left(M,R_{\boldsymbol{\theta}}(M)\right):
\dfrac{\mathrm dM_{\boldsymbol{\theta}}}{\mathrm d\epsilon_c}>0\right\},
& d\in\mathrm{NICER},\\[2mm]
\left\{\left(M,\ln\Lambda_{\boldsymbol{\theta}}(M)\right):
\dfrac{\mathrm dM_{\boldsymbol{\theta}}}{\mathrm d\epsilon_c}>0\right\},
& d\in\mathrm{GW},
\end{cases}
\end{equation}
denotes the stable two-dimensional stellar sequence predicted by the EoS parameters \(\boldsymbol{\theta}\) in the observable plane associated with data set \(d\).

\begin{table*}[t]
\centering
\caption{Parameters of the bivariate-Gaussian approximations to the six two-dimensional observational constraints. For the NICER data sets, \((x_d,y_d)=(M,R)\), whereas for the gravitational-wave events, \((x_d,y_d)=(M,\ln\Lambda)\). The quantities \(\mu_{x,d}\) and \(\sigma_{x,d}\) are given in \(M_\odot\); \(\mu_{y,d}\) and \(\sigma_{y,d}\) are given in km for the NICER measurements and are dimensionless for the gravitational-wave events.}
\label{tab:2d_gaussian_parameters}
\begin{tabular}{lcccccc}
\hline
Data set & \((x_d,y_d)\) & \(\mu_{x,d}\) & \(\mu_{y,d}\) & \(\sigma_{x,d}\) & \(\sigma_{y,d}\) & \(\rho_d\) \\
\hline
PSR J0030$+$0451 & \((M,R)\) & 1.4531 & 12.7899 & 0.1901 & 1.1778 & 0.7724 \\
PSR J0437$-$4715 & \((M,R)\) & 1.4179 & 11.4806 & 0.03477 & 0.7891 & 0.2608 \\
PSR J0614$-$3329 & \((M,R)\) & 1.4440 & 10.3594 & 0.06432 & 0.9327 & 0.3752 \\
PSR J0740$+$6620 & \((M,R)\) & 2.0730 & 12.6519 & 0.06873 & 1.0811 & 0.2710 \\
GW170817 & \((M,\ln\Lambda)\) & 1.3701 & 5.6684 & 0.1207 & 1.1972 & $-0.2374$ \\
GW190425 & \((M,\ln\Lambda)\) & 1.6563 & 5.7102 & 0.1267 & 1.2967 & $-0.1770$ \\
\hline
\end{tabular}
\end{table*}

In this work, the independent mass measurements of PSRs J0348$+$0432 and J1614$-$2230 \cite{Antoniadis:2013pzd,NANOGRAV:2018hou} are treated as one-dimensional Gaussian constraints, with \((\mu_{M,a},\sigma_{M,a})=(2.01,0.04)\,M_\odot\) for J0348$+$0432 and \((1.908,0.016)\,M_\odot\) for J1614$-$2230. For each pulsar \(a\), the peak-normalized observational likelihood is
\begin{equation}
\mathcal L_{a,\mathrm{obs}}^{M}(M)=\exp\left[-\frac{1}{2}\left(\frac{M-\mu_{M,a}}{\sigma_{M,a}}\right)^2\right],\qquad \mathcal L_{a,\mathrm{obs}}^{M}(\mu_{M,a})=1.
\label{eq:massive_pulsar_gaussian}
\end{equation}
And, in each EoS, the likelihood parameter set \(\boldsymbol{\theta}\) is obtained by maximizing this observational likelihood over the mass coordinate of its stable mass--radius sequence,
\begin{equation}
\mathcal L_a^{\rm 1D}(\boldsymbol{\theta})=\operatorname*{max}_{(M,R)\in\mathcal S_{\boldsymbol{\theta}}^{MR}}\mathcal L_{a,\mathrm{obs}}^{M}(M).
\label{eq:massive_pulsar_profile}
\end{equation}

The nuclear-matter likelihood is factorized as
\begin{equation}
\mathcal L_{\rm NM}(\boldsymbol{\theta})=\prod_{b\in\{n_0,E_0,K_0,J,L\}}\mathcal L_b^{\rm NM}(\boldsymbol{\theta})=\exp\left[-\frac{1}{2}\sum_{b\in\{n_0,E_0,K_0,J,L\}}\left(\frac{b(\boldsymbol{\theta})-\mu_b}{\sigma_b}\right)^2\right],
\label{eq:NM_likelihood}
\end{equation}
where \(b(\boldsymbol{\theta})\) is the nuclear-matter quantity predicted by the EoS parameters \(\boldsymbol{\theta}\), and \(\mu_b\) and \(\sigma_b\) are its empirical central value and standard deviation, respectively. The empirical central values \(\mu_b\) and standard deviations \(\sigma_b\) are listed in Eq.~\eqref{eq:empirical_parameter_priors}.
We assume uniform priors within the adopted parameter ranges, treat the different measurements as independent. For the Taylor metamodel and explicit hadronic models, the five nuclear-matter quantities \(n_0\), \(E_0\), \(K_0\), \(J\), and \(L\) are included together with the astronomical constraints stisfying  
\begin{equation}
T(\boldsymbol{\theta})
\equiv-2\ln\!\left[\mathcal L_{\rm NM}(\boldsymbol{\theta})\mathcal L_{\rm astro}(\boldsymbol{\theta})\right]
\leq\chi_{13}^2(0.6827)=14.8421.
\label{eq:hadronic_joint_1sigma}
\end{equation}
The NN and SR EoSs instead describe beta-equilibrated matter
without separately determining these five nuclear-matter quantities. They are therefore constrained only by the astronomical data through
\begin{equation}
T_{\beta{\rm eq}}(\boldsymbol{\theta})
\equiv-2\ln\mathcal L_{\rm astro}(\boldsymbol{\theta})
\leq\chi_8^2(0.6827)=9.3039.
\label{eq:beta_equilibrated_joint_1sigma}
\end{equation}
These conditions provide practical definitions of the joint-\(1\sigma\) allowed regions under the stated assumptions, rather than exact Bayesian \(68.27\%\) credible regions.

\section{Nuclear matter parametrization and EoS construction}
\label{app:EoS_parametrization}

This Appendix describes the two EoS parametrizations used in our analysis, the Taylor nuclear metamodel employed to construct the nucleonic reference domain and the NN--SR framework used to generate an independent ensemble of smooth EoSs.

\subsection{Taylor metamodel around saturation density}
\label{app:EoS_parametrization_metamodel}

We describe homogeneous cold nuclear matter in terms of the baryon density $n=n_n+n_p$ and the isospin asymmetry
\begin{equation}
 \delta=\frac{n_n-n_p}{n}=1-2Y_p,
 \qquad Y_p=\frac{n_p}{n},
\end{equation}
with $Y_p$ being the portion of the protons in the nuclear matter and $n_{n(p)}$ being the the density or neutron(proton).  Within the parabolic approximation, the energy per-baryon excluding the nucleon rest-mass is written as
\begin{equation}
 E(n,\delta)=E_{\rm SNM}(n)+S(n)\delta^2,
 \label{eq:parabolic_nuclear_energy}
\end{equation}
where $E_{\rm SNM}$ is the energy of symmetric nuclear matter and $S(n)$ is the symmetry energy. Both functions are expanded around the saturation density $n_0$ using the dimensionless variable
\begin{equation}
 x=\frac{n-n_0}{3n_0}
   =\frac{1}{3}\left(\frac{n}{n_0}-1\right).
 \label{eq:taylor_variable}
\end{equation}
The factor $3n_0$ makes the expansion coefficients coincide with the standard nuclear matter empirical parameters.

Up to sixth order of $x$, the energy of symmetric nuclear matter $ E_{\rm SNM}(n)$ is expressed as
\begin{align}
 E_{\rm SNM}(n)={}&E_0
 +\frac{1}{2!}K_0x^2
 +\frac{1}{3!}Q_0x^3
 +\frac{1}{4!}Z_0x^4
 +\frac{1}{5!}X_0x^5
 +\frac{1}{6!}Y_0x^6,
 \label{eq:snm_taylor_expansion}
\end{align}
where the term linear in $x$ vanishes because symmetric matter has zero pressure at saturation density. The symmetry energy $S(n)$ is
\begin{align}
 S(n)={}&J+Lx
 +\frac{1}{2!}K_{\rm sym}x^2
 +\frac{1}{3!}Q_{\rm sym}x^3
 +\frac{1}{4!}Z_{\rm sym}x^4
 +\frac{1}{5!}X_{\rm sym}x^5
 +\frac{1}{6!}Y_{\rm sym}x^6.
 \label{eq:symmetry_taylor_expansion}
\end{align}
The coefficients are defined by derivatives evaluated at $n_0$:
\begin{align}
 E_0&=E_{\rm SNM}(n_0),
 &K_0&=9n_0^2E_{\rm SNM}^{(2)}(n_0),
 &Q_0&=27n_0^3E_{\rm SNM}^{(3)}(n_0), \nonumber\\
 Z_0&=81n_0^4E_{\rm SNM}^{(4)}(n_0),
 &X_0&=243n_0^5E_{\rm SNM}^{(5)}(n_0),
 &Y_0&=729n_0^6E_{\rm SNM}^{(6)}(n_0),
 \label{eq:isoscalar_derivative_definitions}
\end{align}
and
\begin{align}
 J&=S(n_0),
 &L&=3n_0S^{(1)}(n_0),
 &K_{\rm sym}&=9n_0^2S^{(2)}(n_0), \nonumber\\
 Q_{\rm sym}&=27n_0^3S^{(3)}(n_0),
 &Z_{\rm sym}&=81n_0^4S^{(4)}(n_0), \nonumber\\
 X_{\rm sym}&=243n_0^5S^{(5)}(n_0),
 &Y_{\rm sym}&=729n_0^6S^{(6)}(n_0),
 \label{eq:isovector_derivative_definitions}
\end{align}
where the upper index $``(i)"$ stands for the $i$-th derivative with respect to density $n$.
Here $K_0$, $Q_0$, and $Z_0$ are the isoscalar incompressibility, skewness, and kurtosis parameters, while $L$, $K_{\rm sym}$, $Q_{\rm sym}$, and $Z_{\rm sym}$ characterize the corresponding density dependence of the symmetry energy. The symbols $X_{0, \rm{sym}}$ and $Y_{0, \rm{sym}}$ denote the fifth- and sixth-order derivative coefficients which will be used to test the convergence of the expansion.

In the main text, the constraints on the EoSs from the multimessenger inference is performed at fourth order of Eqs.~\eqref{eq:snm_taylor_expansion} and \eqref{eq:symmetry_taylor_expansion}, for which $X_0=X_{\rm sym}=Y_0=Y_{\rm sym}=0$. To test whether the inferred functional region is stable against truncation, we additionally repeat the calculation at orders $N=2,3,4,5$, and $6$. At a given order $N$, coefficients of higher powers than $x^N$ are set to zero, and the lower-order parameters and all physical and astronomical filters are treated in the same manner.

The empirical saturation parameters are taken as~\cite{Dutra:2014qga,Lattimer:2012xj,Danielewicz:2013upa,Dutra:2012mb}:
\begin{align}
 n_0&=0.160\pm0.005\ {\rm fm}^{-3},
 &E_0&=-16.0\pm1.0\ {\rm MeV},
\nonumber\\
 J&=30.9\pm1.9\ {\rm MeV},
 &L&=52.5\pm17.5\ {\rm MeV},
 &K_0&=230\pm30\ {\rm MeV}.
 \label{eq:empirical_parameter_priors}
\end{align}
In our analysis, the quoted uncertainties are used as empirical reference scales rather than being assigned a specific confidence level. To allow a broader exploration of the parameter space, the half-width of each sampling interval is enlarged by a factor of four relative to the corresponding quoted spread. 

For \(Q_0\), Cai {\it et al}. obtained \([-494,-10]\) MeV from a nonlinear relativistic mean-field analysis constrained by heavy-ion flow data and massive NSs~\cite{Cai:2014kya}, while the Bayesian heavy-ion analysis of Xie and Li yielded \(Q_0=-180^{+100}_{-110}\) MeV~\cite{Xie:2020kta}. Margueron {\it et al}.~\cite{Margueron:2017eqc} compiled empirical parameters from experimental analyses and a large set of relativistic and non-relativistic nuclear interactions, obtaining \(Q_0=300\pm400\), \(Z_0=-500\pm1000\), \(K_{\rm sym}=-100\pm100\), \(Q_{\rm sym}=0\pm400\), and \(Z_{\rm sym}=-500\pm1000\) MeV. To accommodate the substantial spread among these estimates, we assign independent uniform priors,
\begin{align}
 Q_0 &= 300\pm800\ {\rm MeV},
 &Z_0 &= -500\pm2000\ {\rm MeV}, \nonumber\\
 K_{\rm sym} &= -100\pm200\ {\rm MeV},
 &Q_{\rm sym} &= 0\pm800\ {\rm MeV}, \nonumber\\
 Z_{\rm sym} &= -500\pm2000\ {\rm MeV}.
 \label{eq:higher_parameter_priors}
\end{align}
For the fifth- and sixth-order convergence calculations, we use typical values $X_0,X_{\rm sym}\in[-5000,5000]$ MeV and $Y_0,Y_{\rm sym}\in[-10000,10000]$ MeV. These broad intervals are sampling domains for the truncation test and should not be interpreted as empirical measurements of the corresponding derivatives.

In this work, the cold, charge-neutral \(npe\) matter is constructed by using the beta equilibrium conditions
\begin{equation}
\mu_n-\mu_p=\mu_e,
\qquad
n_p=n_e,
\label{eq:beta_equilibrium}
\end{equation}
where \(\delta=1-2Y_p\) and \(Y_p=n_p/n\). Electrons are treated as an ideal relativistic Fermi gas. Equation~\eqref{eq:beta_equilibrium} is solved at each density to determine \(Y_p(n)\). In addition, the uniform-matter EoS is connected to the BPS~\cite{Baym:1971pw} crust through a thermodynamically consistent Maxwell construction. For each EoS, the transition densities \(n_-\) on the BPS branch and \(n_+\) on the uniform-matter branch are determined by the coexistence conditions
\begin{equation}
P_{\rm BPS}(n_-)=P_{\rm core}(n_+)=P_t,\qquad
\mu_{\rm BPS}(n_-)=\mu_{\rm core}(n_+)=\mu_t,
\end{equation}
where \(\mu=(\varepsilon+P)/n\) is the baryon chemical potential. The resulting stellar EoS follows the BPS branch below \(n_-\), crosses the crust--core coexistence interval through a density discontinuity from \(n_-\) to \(n_+\) at constant pressure \(P_t\), and follows the continuous uniform-matter branch above \(n_+\). The associated energy-density discontinuity is
\begin{equation}
\Delta\varepsilon_t=\varepsilon_{\rm core}(n_+)-\varepsilon_{\rm BPS}(n_-)
=\mu_t(n_+-n_-).
\end{equation}
The uniform-matter branch is evaluated over \(n_+\leq n\leq8n_0\).  We require that throughout this interval:
\begin{equation}
 \begin{aligned}
 S(n)&>0, &\varepsilon(n)&>0, &P(n)&>0,\\
 \frac{d\varepsilon}{dn}&\geq0,
 &\frac{dP}{dn}&\geq0,
 &0<c_s^2&\leq1,
 &0\leq Y_p&\leq\frac12.
 \end{aligned}
 \label{eq:EoS_physical_filters}
\end{equation}

As shown in Fig.~\ref{fig:gp}, the inferred thermodynamic region changes less as the expansion order increases and is nearly unchanged beyond fourth order. The fourth-order expansion is therefore adopted for the main analysis. 

\subsection{Neural-network reconstruction and symbolic regression of the equation of state}
\label{app:EoS_parametrization_NN}

Combining NNs with SR provides complementary advantages for EoS reconstruction. NNs offer the flexibility to represent a broad class of nonlinear barotropic relations without imposing a fixed functional form. SR then distils each numerical reconstruction into a compact analytic expression, providing functional interpretability, exposing the dominant mathematical dependence and allowing thermodynamic derivatives to be evaluated directly and continuously. The former is therefore used to explore the space of astrophysically admissible EoSs, while the latter makes their thermodynamic behaviour transparent and analytically tractable. This interpretability concerns the functional form of the EoS and does not by itself assign a microscopic composition.

To minimize the model dependence introduced by prescribing a specific high-density functional form, we performed large-scale sampling directly in the space of SV functions using NNs. The SV was represented as
\begin{equation}
c_s^2(\rho)=\rho\,{\rm NN}_{\boldsymbol{\theta}}(\rho),\qquad \rho=\frac{n}{n_0},
\label{eq:neural_cs2}
\end{equation}
where ${\rm NN}_{\boldsymbol{\theta}}$ is a fully connected $1$--$16$--$16$--$1$ network with ReLU activation functions in the two hidden layers and a linear output layer, containing 321 trainable weights and biases. To ensure that the inferred EoS satisfies both physical principles and astrophysical observational constraints, a composite loss function is constructed during the training procedure. The total loss function is written as
\begin{equation}
Loss_{\mathrm{total}}
=Loss_{phy}+ Loss_{\mathrm{MR}}.
\label{eq:Losstotal}
\end{equation}
On a density grid \(\{\rho_i\}_{i=1}^{N}\), the physical penalty is defined as:
\be
    Loss_{\mathrm{phy}} & = & \mathrm{ReLU}(-c_s^2)^2+\mathrm{ReLU}(c_s^2-1)^2+(c_s^2(0))^2 \nonumber\\
    & &{} +\mathrm{ReLU}\left(-(P_{i+1}-P_i)\right)^2 \nonumber\\
    & &{} +\sum_i\mathrm{ReLU}(-P_i)+\sum_i\mathrm{ReLU}(-\varepsilon_i),
    \label{eq:losscc}
\ee
where \({\rm ReLU}(x)=\max(0,x)\). These terms impose the boundary condition, thermodynamic stability, causality, pressure monotonicity, and positivity of the reconstructed EOS. The term \({\rm ReLU}(-c_s^2)^2\) penalizes negative SV and therefore enforces thermodynamic stability, whereas \({\rm ReLU}(c_s^2-1)^2\) penalizes superluminal propagation and enforces causality. The term \([c_s^2(0)]^2\) imposes the boundary condition \(c_s^2(0)=0\). The finite-difference term \({\rm ReLU}[-(P_{i+1}-P_i)]^2\) penalizes any decrease in pressure between successive density-grid points. Finally, \(\sum_i{\rm ReLU}(-P_i)\) and \(\sum_i{\rm ReLU}(-\varepsilon_i)\) penalize negative pressure and negative energy density, respectively.

In Eq.~\eqref{eq:losscc}, $Loss_{\mathrm{MR}}$ represents the constraints from atronomical data which shown in Tab.~\ref{tab:observational_data_sources}.
It is defined by: 
\begin{equation}
Loss_{\mathrm{MR}} = -\mathcal L_{\rm astro}(\boldsymbol{\theta}),
\label{eq:astronomical_objective}
\end{equation}
where \(\boldsymbol{\theta}\) denotes the NN parameters. The construction of the joint astronomical likelihood \((\mathcal L_{\rm astro})\) is detailed in Appendix.~\ref{app:multimessenger_likelihood}. 
The final loss function therefore combines physical consistency conditions and astrophysical observational constraints into a unified optimization framework, enabling the NN to infer physically admissible EoSs that remain consistent with NS observations. The factor $\rho$ was introduced only to enforce $c_s^2(0)=0$. The network was defined over \(0\leq\rho\leq8\), with
\(\rho=n/n_0\), and provides the SV profile
\(c_s^2(n)={\rm d}P/{\rm d}\varepsilon\). Starting from this profile, we reconstruct the EoS by integrating
\begin{equation}
\frac{{\rm d}\varepsilon}{{\rm d}n}
=\frac{\varepsilon+P}{n},
\qquad
\frac{{\rm d}P}{{\rm d}n}
=c_s^2(n)\frac{\varepsilon+P}{n}.
\end{equation}
The integration constants are fixed at \(n_{\rm match}=0.5n_0\) by the BPS values of \(P\) and \(\varepsilon\). The reconstructed branch is used over \(0.5\leq n/n_0\leq8\) and is combined with the BPS EoS~\cite{Baym:1971pw} below \(n_{\rm match}\) to obtain the complete relation \(P(\varepsilon)\). 
We therefore approximate the crust--core transition by the zero-density-jump limit of Maxwell matching, \(n_-=n_+=n_{\rm match}\). For each network parameter vector, the thermodynamic relations were integrated to reconstruct $P(\varepsilon)$, after which the coupled TOV--Love equations were solved to obtain the mass--radius relation, maximum mass, and tidal deformability. Only single-branch continuous EoSs satisfying $0<c_s^2<1$, monotonically increasing pressure and energy density, the existence of a stable stellar branch, and a maximum-mass turning point below $8n_0$ were retained.

The large-scale search was distributed over nine independent evolutionary islands running on three CPU servers. Each island evaluated 4,800 NNs per-generation for 40 generations, corresponding to a planned total of $1.728\times10^6$ candidate networks. Because the full TOV--Love calculation is not differentiable, the network parameters were optimized using derivative-free elite evolution rather than gradient backpropagation. At each generation, high-likelihood and functionally diverse candidates were retained and locally mutated, while $15\%$ of the population was replaced by newly randomized networks to maintain broad exploration of the SV function space. Candidates were ranked using the joint astronomical likelihood constructed from the NICER mass--radius measurements, massive-pulsar masses, and the tidal-deformability information listed in Tab.~\ref{tab:observational_data_sources}.

The accepted NN EoSs are initially available as numerical functions. We therefore apply SR to obtain an analytic representation of each relation. 
SR is a data-driven method for searching analytical expressions that approximate a given set of numerical data. 
Unlike conventional regression, where the functional form is fixed in advance and only the parameters are fitted, SR searches both the functional structure and the numerical coefficients. 
Starting from a predefined library of elementary operators, such as \(+,-,\times,\div,\exp,\log\), and power functions, SR constructs candidate formulas and ranks them according to their accuracy and complexity. 
In this sense, SR provides a useful bridge between flexible numerical representations and compact analytical models.
In SR, the complexity of an expression is usually quantified by assigning a cost to each elementary building block. 
A simple linear expression has low complexity because it contains only constants, variables, and basic arithmetic operations. 
Expressions involving nonlinear functions, such as powers, exponentials, logarithms, or nested structures, are assigned larger complexity costs. 
In this work, we use the complexity assignment summarized in Tab.~\ref{tab:SRcomplexity}.

The complexity measure serves as a regularization criterion in SR. 
For two candidate expressions with comparable fitting accuracy, the expression with smaller complexity is preferred, because it is easier to interpret and less likely to overfit the numerical data. 
Conversely, a higher-complexity expression is accepted only when the improvement in accuracy is sufficient to justify the additional symbolic structure.

In the present work, SR is applied to the SV profile learned by the NN. 
After the NN generates physically admissible EoSs satisfying the imposed NS constraints, the corresponding numerical relation between the baryon density and the SV,
\[
\rho_B \longrightarrow c_s^2(\rho_B),
\]
is used as the target data for SR. 
The goal is to find a compact analytical expression \(c_{s,{\rm SR}}^2(\rho_B)\) that reproduces the NN output while keeping the formula sufficiently simple for physical interpretation. 
The quality of a candidate expression is measured by the deviation between \(c_{s,{\rm SR}}^2(\rho_B)\) and the NN result, together with a penalty for excessive symbolic complexity.

\begin{table}[t]
\centering
\caption{Complexity assignment used in symbolic regression. 
The total complexity of a candidate expression is obtained by summing the costs of all operators, variables, and numerical constants appearing in the expression.}
\label{tab:SRcomplexity}
\begin{tabular}{lc}
\hline
Symbolic component & Complexity cost \\
\hline
Variable, e.g., \(x\) or \(\rho_B\) & 1 \\
Numerical constant & 1 \\
Addition or subtraction, \(+,-\) & 1 \\
Multiplication or division, \(\times,\div\) & 1 \\
Power function, \(x^\alpha\) & 2 \\
Exponential or logarithm, \(\exp,\log\) & 3 \\
Trigonometric functions, \(\sin,\cos\) & 3 \\
Nested nonlinear operation & Sum of all internal costs \\
\hline
\end{tabular}
\end{table}

\end{widetext}

\bibliography{NSEoSNNRef}

\begin{thebibliography}{55}%
\makeatletter
\providecommand \@ifxundefined [1]{%
 \@ifx{#1\undefined}
}%
\providecommand \@ifnum [1]{%
 \ifnum #1\expandafter \@firstoftwo
 \else \expandafter \@secondoftwo
 \fi
}%
\providecommand \@ifx [1]{%
 \ifx #1\expandafter \@firstoftwo
 \else \expandafter \@secondoftwo
 \fi
}%
\providecommand \natexlab [1]{#1}%
\providecommand \enquote  [1]{``#1''}%
\providecommand \bibnamefont  [1]{#1}%
\providecommand \bibfnamefont [1]{#1}%
\providecommand \citenamefont [1]{#1}%
\providecommand \href@noop [0]{\@secondoftwo}%
\providecommand \href [0]{\begingroup \@sanitize@url \@href}%
\providecommand \@href[1]{\@@startlink{#1}\@@href}%
\providecommand \@@href[1]{\endgroup#1\@@endlink}%
\providecommand \@sanitize@url [0]{\catcode `\\12\catcode `\$12\catcode
  `\&12\catcode `\#12\catcode `\^12\catcode `\_12\catcode `\%12\relax}%
\providecommand \@@startlink[1]{}%
\providecommand \@@endlink[0]{}%
\providecommand \url  [0]{\begingroup\@sanitize@url \@url }%
\providecommand \@url [1]{\endgroup\@href {#1}{\urlprefix }}%
\providecommand \urlprefix  [0]{URL }%
\providecommand \Eprint [0]{\href }%
\providecommand \doibase [0]{http://dx.doi.org/}%
\providecommand \selectlanguage [0]{\@gobble}%
\providecommand \bibinfo  [0]{\@secondoftwo}%
\providecommand \bibfield  [0]{\@secondoftwo}%
\providecommand \translation [1]{[#1]}%
\providecommand \BibitemOpen [0]{}%
\providecommand \bibitemStop [0]{}%
\providecommand \bibitemNoStop [0]{.\EOS\space}%
\providecommand \EOS [0]{\spacefactor3000\relax}%
\providecommand \BibitemShut  [1]{\csname bibitem#1\endcsname}%
\let\auto@bib@innerbib\@empty
\bibitem [{\citenamefont {Antoniadis}\ \emph {et~al.}(2013)\citenamefont
  {Antoniadis} \emph {et~al.}}]{Antoniadis:2013pzd}%
  \BibitemOpen
  \bibfield  {author} {\bibinfo {author} {\bibfnamefont {J.}~\bibnamefont
  {Antoniadis}} \emph {et~al.},\ }\href {\doibase 10.1126/science.1233232}
  {\bibfield  {journal} {\bibinfo  {journal} {Science}\ }\textbf {\bibinfo
  {volume} {340}},\ \bibinfo {pages} {6131} (\bibinfo {year} {2013})},\ \Eprint
  {http://arxiv.org/abs/1304.6875} {arXiv:1304.6875 [astro-ph.HE]} \BibitemShut
  {NoStop}%
\bibitem [{\citenamefont {Abbott}\ \emph {et~al.}(2020)\citenamefont {Abbott}
  \emph {et~al.}}]{LIGOScientific:2020aai}%
  \BibitemOpen
  \bibfield  {author} {\bibinfo {author} {\bibfnamefont {B.~P.}\ \bibnamefont
  {Abbott}} \emph {et~al.} (\bibinfo {collaboration} {LIGO Scientific,
  Virgo}),\ }\href {\doibase 10.3847/2041-8213/ab75f5} {\bibfield  {journal}
  {\bibinfo  {journal} {Astrophys. J. Lett.}\ }\textbf {\bibinfo {volume}
  {892}},\ \bibinfo {pages} {L3} (\bibinfo {year} {2020})},\ \Eprint
  {http://arxiv.org/abs/2001.01761} {arXiv:2001.01761 [astro-ph.HE]}
  \BibitemShut {NoStop}%
\bibitem [{\citenamefont {Abbott}\ \emph {et~al.}(2019)\citenamefont {Abbott}
  \emph {et~al.}}]{LIGOScientific:2018hze}%
  \BibitemOpen
  \bibfield  {author} {\bibinfo {author} {\bibfnamefont {B.~P.}\ \bibnamefont
  {Abbott}} \emph {et~al.} (\bibinfo {collaboration} {LIGO Scientific,
  Virgo}),\ }\href {\doibase 10.1103/PhysRevX.9.011001} {\bibfield  {journal}
  {\bibinfo  {journal} {Phys. Rev. X}\ }\textbf {\bibinfo {volume} {9}},\
  \bibinfo {pages} {011001} (\bibinfo {year} {2019})},\ \Eprint
  {http://arxiv.org/abs/1805.11579} {arXiv:1805.11579 [gr-qc]} \BibitemShut
  {NoStop}%
\bibitem [{\citenamefont {Choudhury}\ \emph {et~al.}(2024)\citenamefont
  {Choudhury} \emph {et~al.}}]{Choudhury:2024xbk}%
  \BibitemOpen
  \bibfield  {author} {\bibinfo {author} {\bibfnamefont {D.}~\bibnamefont
  {Choudhury}} \emph {et~al.},\ }\href {\doibase 10.3847/2041-8213/ad5a6f}
  {\bibfield  {journal} {\bibinfo  {journal} {Astrophys. J. Lett.}\ }\textbf
  {\bibinfo {volume} {971}},\ \bibinfo {pages} {L20} (\bibinfo {year}
  {2024})},\ \Eprint {http://arxiv.org/abs/2407.06789} {arXiv:2407.06789
  [astro-ph.HE]} \BibitemShut {NoStop}%
\bibitem [{\citenamefont {Baym}\ \emph {et~al.}(2018)\citenamefont {Baym},
  \citenamefont {Hatsuda}, \citenamefont {Kojo}, \citenamefont {Powell},
  \citenamefont {Song},\ and\ \citenamefont {Takatsuka}}]{Baym:2017whm}%
  \BibitemOpen
  \bibfield  {author} {\bibinfo {author} {\bibfnamefont {G.}~\bibnamefont
  {Baym}}, \bibinfo {author} {\bibfnamefont {T.}~\bibnamefont {Hatsuda}},
  \bibinfo {author} {\bibfnamefont {T.}~\bibnamefont {Kojo}}, \bibinfo {author}
  {\bibfnamefont {P.~D.}\ \bibnamefont {Powell}}, \bibinfo {author}
  {\bibfnamefont {Y.}~\bibnamefont {Song}}, \ and\ \bibinfo {author}
  {\bibfnamefont {T.}~\bibnamefont {Takatsuka}},\ }\href {\doibase
  10.1088/1361-6633/aaae14} {\bibfield  {journal} {\bibinfo  {journal} {Rept.
  Prog. Phys.}\ }\textbf {\bibinfo {volume} {81}},\ \bibinfo {pages} {056902}
  (\bibinfo {year} {2018})},\ \Eprint {http://arxiv.org/abs/1707.04966}
  {arXiv:1707.04966 [astro-ph.HE]} \BibitemShut {NoStop}%
\bibitem [{\citenamefont {Chen}(2017)}]{Chen:2017och}%
  \BibitemOpen
  \bibfield  {author} {\bibinfo {author} {\bibfnamefont {L.-W.}\ \bibnamefont
  {Chen}},\ }\href {\doibase 10.11804/NuclPhysRev.34.01.020} {\bibfield
  {journal} {\bibinfo  {journal} {Nucl. Phys. Rev.}\ }\textbf {\bibinfo
  {volume} {34}},\ \bibinfo {pages} {20} (\bibinfo {year} {2017})},\ \Eprint
  {http://arxiv.org/abs/1708.04433} {arXiv:1708.04433 [nucl-th]} \BibitemShut
  {NoStop}%
\bibitem [{\citenamefont {Gil}\ \emph {et~al.}(2019)\citenamefont {Gil},
  \citenamefont {Papakonstantinou}, \citenamefont {Hyun},\ and\ \citenamefont
  {Oh}}]{Gil:2018yah}%
  \BibitemOpen
  \bibfield  {author} {\bibinfo {author} {\bibfnamefont {H.}~\bibnamefont
  {Gil}}, \bibinfo {author} {\bibfnamefont {P.}~\bibnamefont
  {Papakonstantinou}}, \bibinfo {author} {\bibfnamefont {C.~H.}\ \bibnamefont
  {Hyun}}, \ and\ \bibinfo {author} {\bibfnamefont {Y.}~\bibnamefont {Oh}},\
  }\href {\doibase 10.1103/PhysRevC.99.064319} {\bibfield  {journal} {\bibinfo
  {journal} {Phys. Rev. C}\ }\textbf {\bibinfo {volume} {99}},\ \bibinfo
  {pages} {064319} (\bibinfo {year} {2019})},\ \Eprint
  {http://arxiv.org/abs/1805.11321} {arXiv:1805.11321 [nucl-th]} \BibitemShut
  {NoStop}%
\bibitem [{\citenamefont {Sorensen}\ \emph {et~al.}(2024)\citenamefont
  {Sorensen} \emph {et~al.}}]{Sorensen:2023zkk}%
  \BibitemOpen
  \bibfield  {author} {\bibinfo {author} {\bibfnamefont {A.}~\bibnamefont
  {Sorensen}} \emph {et~al.},\ }\href {\doibase 10.1016/j.ppnp.2023.104080}
  {\bibfield  {journal} {\bibinfo  {journal} {Prog. Part. Nucl. Phys.}\
  }\textbf {\bibinfo {volume} {134}},\ \bibinfo {pages} {104080} (\bibinfo
  {year} {2024})},\ \Eprint {http://arxiv.org/abs/2301.13253} {arXiv:2301.13253
  [nucl-th]} \BibitemShut {NoStop}%
\bibitem [{\citenamefont {Horowitz}\ \emph {et~al.}(2014)\citenamefont
  {Horowitz}, \citenamefont {Brown}, \citenamefont {Kim}, \citenamefont
  {Lynch}, \citenamefont {Michaels}, \citenamefont {Ono}, \citenamefont
  {Piekarewicz}, \citenamefont {Tsang},\ and\ \citenamefont
  {Wolter}}]{Horowitz:2014bja}%
  \BibitemOpen
  \bibfield  {author} {\bibinfo {author} {\bibfnamefont {C.~J.}\ \bibnamefont
  {Horowitz}}, \bibinfo {author} {\bibfnamefont {E.~F.}\ \bibnamefont {Brown}},
  \bibinfo {author} {\bibfnamefont {Y.}~\bibnamefont {Kim}}, \bibinfo {author}
  {\bibfnamefont {W.~G.}\ \bibnamefont {Lynch}}, \bibinfo {author}
  {\bibfnamefont {R.}~\bibnamefont {Michaels}}, \bibinfo {author}
  {\bibfnamefont {A.}~\bibnamefont {Ono}}, \bibinfo {author} {\bibfnamefont
  {J.}~\bibnamefont {Piekarewicz}}, \bibinfo {author} {\bibfnamefont {M.~B.}\
  \bibnamefont {Tsang}}, \ and\ \bibinfo {author} {\bibfnamefont {H.~H.}\
  \bibnamefont {Wolter}},\ }\href {\doibase 10.1088/0954-3899/41/9/093001}
  {\bibfield  {journal} {\bibinfo  {journal} {J. Phys. G}\ }\textbf {\bibinfo
  {volume} {41}},\ \bibinfo {pages} {093001} (\bibinfo {year} {2014})},\
  \Eprint {http://arxiv.org/abs/1401.5839} {arXiv:1401.5839 [nucl-th]}
  \BibitemShut {NoStop}%
\bibitem [{\citenamefont {Demorest}\ \emph {et~al.}(2010)\citenamefont
  {Demorest}, \citenamefont {Pennucci}, \citenamefont {Ransom}, \citenamefont
  {Roberts},\ and\ \citenamefont {Hessels}}]{Demorest:2010bx}%
  \BibitemOpen
  \bibfield  {author} {\bibinfo {author} {\bibfnamefont {P.}~\bibnamefont
  {Demorest}}, \bibinfo {author} {\bibfnamefont {T.}~\bibnamefont {Pennucci}},
  \bibinfo {author} {\bibfnamefont {S.}~\bibnamefont {Ransom}}, \bibinfo
  {author} {\bibfnamefont {M.}~\bibnamefont {Roberts}}, \ and\ \bibinfo
  {author} {\bibfnamefont {J.}~\bibnamefont {Hessels}},\ }\href {\doibase
  10.1038/nature09466} {\bibfield  {journal} {\bibinfo  {journal} {Nature}\
  }\textbf {\bibinfo {volume} {467}},\ \bibinfo {pages} {1081} (\bibinfo {year}
  {2010})}\BibitemShut {NoStop}%
\bibitem [{\citenamefont {Vinciguerra}\ \emph {et~al.}(2024)\citenamefont
  {Vinciguerra} \emph {et~al.}}]{Vinciguerra:2023qxq}%
  \BibitemOpen
  \bibfield  {author} {\bibinfo {author} {\bibfnamefont {S.}~\bibnamefont
  {Vinciguerra}} \emph {et~al.},\ }\href {\doibase 10.3847/1538-4357/acfb83}
  {\bibfield  {journal} {\bibinfo  {journal} {Astrophys. J.}\ }\textbf
  {\bibinfo {volume} {961}},\ \bibinfo {pages} {62} (\bibinfo {year} {2024})},\
  \Eprint {http://arxiv.org/abs/2308.09469} {arXiv:2308.09469 [astro-ph.HE]}
  \BibitemShut {NoStop}%
\bibitem [{\citenamefont {Mauviard}\ \emph {et~al.}(2025)\citenamefont
  {Mauviard} \emph {et~al.}}]{Mauviard:2025dmd}%
  \BibitemOpen
  \bibfield  {author} {\bibinfo {author} {\bibfnamefont {L.}~\bibnamefont
  {Mauviard}} \emph {et~al.},\ }\href {\doibase 10.3847/1538-4357/ae145d}
  {\bibfield  {journal} {\bibinfo  {journal} {Astrophys. J.}\ }\textbf
  {\bibinfo {volume} {995}},\ \bibinfo {pages} {60} (\bibinfo {year} {2025})},\
  \Eprint {http://arxiv.org/abs/2506.14883} {arXiv:2506.14883 [astro-ph.HE]}
  \BibitemShut {NoStop}%
\bibitem [{\citenamefont {Altiparmak}\ \emph {et~al.}(2022)\citenamefont
  {Altiparmak}, \citenamefont {Ecker},\ and\ \citenamefont
  {Rezzolla}}]{Altiparmak:2022bke}%
  \BibitemOpen
  \bibfield  {author} {\bibinfo {author} {\bibfnamefont {S.}~\bibnamefont
  {Altiparmak}}, \bibinfo {author} {\bibfnamefont {C.}~\bibnamefont {Ecker}}, \
  and\ \bibinfo {author} {\bibfnamefont {L.}~\bibnamefont {Rezzolla}},\ }\href
  {\doibase 10.3847/2041-8213/ac9b2a} {\bibfield  {journal} {\bibinfo
  {journal} {Astrophys. J. Lett.}\ }\textbf {\bibinfo {volume} {939}},\
  \bibinfo {pages} {L34} (\bibinfo {year} {2022})},\ \Eprint
  {http://arxiv.org/abs/2203.14974} {arXiv:2203.14974 [astro-ph.HE]}
  \BibitemShut {NoStop}%
\bibitem [{\citenamefont {Marczenko}\ \emph {et~al.}(2023)\citenamefont
  {Marczenko}, \citenamefont {McLerran}, \citenamefont {Redlich},\ and\
  \citenamefont {Sasaki}}]{Marczenko:2022jhl}%
  \BibitemOpen
  \bibfield  {author} {\bibinfo {author} {\bibfnamefont {M.}~\bibnamefont
  {Marczenko}}, \bibinfo {author} {\bibfnamefont {L.}~\bibnamefont {McLerran}},
  \bibinfo {author} {\bibfnamefont {K.}~\bibnamefont {Redlich}}, \ and\
  \bibinfo {author} {\bibfnamefont {C.}~\bibnamefont {Sasaki}},\ }\href
  {\doibase 10.1103/PhysRevC.107.025802} {\bibfield  {journal} {\bibinfo
  {journal} {Phys. Rev. C}\ }\textbf {\bibinfo {volume} {107}},\ \bibinfo
  {pages} {025802} (\bibinfo {year} {2023})},\ \Eprint
  {http://arxiv.org/abs/2207.13059} {arXiv:2207.13059 [nucl-th]} \BibitemShut
  {NoStop}%
\bibitem [{\citenamefont {Abbott}\ \emph
  {et~al.}(2017{\natexlab{a}})\citenamefont {Abbott} \emph
  {et~al.}}]{LIGOScientific:2017vwq}%
  \BibitemOpen
  \bibfield  {author} {\bibinfo {author} {\bibfnamefont {B.~P.}\ \bibnamefont
  {Abbott}} \emph {et~al.} (\bibinfo {collaboration} {LIGO Scientific,
  Virgo}),\ }\href {\doibase 10.1103/PhysRevLett.119.161101} {\bibfield
  {journal} {\bibinfo  {journal} {Phys. Rev. Lett.}\ }\textbf {\bibinfo
  {volume} {119}},\ \bibinfo {pages} {161101} (\bibinfo {year}
  {2017}{\natexlab{a}})},\ \Eprint {http://arxiv.org/abs/1710.05832}
  {arXiv:1710.05832 [gr-qc]} \BibitemShut {NoStop}%
\bibitem [{\citenamefont {Abbott}\ \emph
  {et~al.}(2017{\natexlab{b}})\citenamefont {Abbott} \emph
  {et~al.}}]{LIGOScientific:2017ync}%
  \BibitemOpen
  \bibfield  {author} {\bibinfo {author} {\bibfnamefont {B.~P.}\ \bibnamefont
  {Abbott}} \emph {et~al.} (\bibinfo {collaboration} {LIGO Scientific, Virgo,
  Fermi GBM, INTEGRAL, IceCube, AstroSat Cadmium Zinc Telluride Imager Team,
  IPN, Insight-Hxmt, ANTARES, Swift, AGILE Team, 1M2H Team, Dark Energy Camera
  GW-EM, DES, DLT40, GRAWITA, Fermi-LAT, ATCA, ASKAP, Las Cumbres Observatory
  Group, OzGrav, DWF (Deeper Wider Faster Program), AST3, CAASTRO, VINROUGE,
  MASTER, J-GEM, GROWTH, JAGWAR, CaltechNRAO, TTU-NRAO, NuSTAR, Pan-STARRS,
  MAXI Team, TZAC Consortium, KU, Nordic Optical Telescope, ePESSTO, GROND,
  Texas Tech University, SALT Group, TOROS, BOOTES, MWA, CALET, IKI-GW
  Follow-up, H.E.S.S., LOFAR, LWA, HAWC, Pierre Auger, ALMA, Euro VLBI Team, Pi
  of Sky, Chandra Team at McGill University, DFN, ATLAS Telescopes, High Time
  Resolution Universe Survey, RIMAS, RATIR, SKA South Africa/MeerKAT}),\ }\href
  {\doibase 10.3847/2041-8213/aa91c9} {\bibfield  {journal} {\bibinfo
  {journal} {Astrophys. J. Lett.}\ }\textbf {\bibinfo {volume} {848}},\
  \bibinfo {pages} {L12} (\bibinfo {year} {2017}{\natexlab{b}})},\ \Eprint
  {http://arxiv.org/abs/1710.05833} {arXiv:1710.05833 [astro-ph.HE]}
  \BibitemShut {NoStop}%
\bibitem [{\citenamefont {Mroczek}\ \emph {et~al.}(2024)\citenamefont
  {Mroczek}, \citenamefont {Miller}, \citenamefont {Noronha-Hostler},\ and\
  \citenamefont {Yunes}}]{Mroczek:2023zxo}%
  \BibitemOpen
  \bibfield  {author} {\bibinfo {author} {\bibfnamefont {D.}~\bibnamefont
  {Mroczek}}, \bibinfo {author} {\bibfnamefont {M.~C.}\ \bibnamefont {Miller}},
  \bibinfo {author} {\bibfnamefont {J.}~\bibnamefont {Noronha-Hostler}}, \ and\
  \bibinfo {author} {\bibfnamefont {N.}~\bibnamefont {Yunes}},\ }\href
  {\doibase 10.1103/PhysRevD.110.123009} {\bibfield  {journal} {\bibinfo
  {journal} {Phys. Rev. D}\ }\textbf {\bibinfo {volume} {110}},\ \bibinfo
  {pages} {123009} (\bibinfo {year} {2024})},\ \Eprint
  {http://arxiv.org/abs/2309.02345} {arXiv:2309.02345 [astro-ph.HE]}
  \BibitemShut {NoStop}%
\bibitem [{\citenamefont {Chatziioannou}\ \emph {et~al.}(2015)\citenamefont
  {Chatziioannou}, \citenamefont {Yagi}, \citenamefont {Klein}, \citenamefont
  {Cornish},\ and\ \citenamefont {Yunes}}]{Chatziioannou:2015uea}%
  \BibitemOpen
  \bibfield  {author} {\bibinfo {author} {\bibfnamefont {K.}~\bibnamefont
  {Chatziioannou}}, \bibinfo {author} {\bibfnamefont {K.}~\bibnamefont {Yagi}},
  \bibinfo {author} {\bibfnamefont {A.}~\bibnamefont {Klein}}, \bibinfo
  {author} {\bibfnamefont {N.}~\bibnamefont {Cornish}}, \ and\ \bibinfo
  {author} {\bibfnamefont {N.}~\bibnamefont {Yunes}},\ }\href {\doibase
  10.1103/PhysRevD.92.104008} {\bibfield  {journal} {\bibinfo  {journal} {Phys.
  Rev. D}\ }\textbf {\bibinfo {volume} {92}},\ \bibinfo {pages} {104008}
  (\bibinfo {year} {2015})},\ \Eprint {http://arxiv.org/abs/1508.02062}
  {arXiv:1508.02062 [gr-qc]} \BibitemShut {NoStop}%
\bibitem [{\citenamefont {Akmal}\ \emph {et~al.}(1998)\citenamefont {Akmal},
  \citenamefont {Pandharipande},\ and\ \citenamefont
  {Ravenhall}}]{Akmal:1998cf}%
  \BibitemOpen
  \bibfield  {author} {\bibinfo {author} {\bibfnamefont {A.}~\bibnamefont
  {Akmal}}, \bibinfo {author} {\bibfnamefont {V.~R.}\ \bibnamefont
  {Pandharipande}}, \ and\ \bibinfo {author} {\bibfnamefont {D.~G.}\
  \bibnamefont {Ravenhall}},\ }\href {\doibase 10.1103/PhysRevC.58.1804}
  {\bibfield  {journal} {\bibinfo  {journal} {Phys. Rev. C}\ }\textbf {\bibinfo
  {volume} {58}},\ \bibinfo {pages} {1804} (\bibinfo {year} {1998})},\ \Eprint
  {http://arxiv.org/abs/nucl-th/9804027} {arXiv:nucl-th/9804027} \BibitemShut
  {NoStop}%
\bibitem [{\citenamefont {Gandolfi}\ \emph {et~al.}(2012)\citenamefont
  {Gandolfi}, \citenamefont {Carlson},\ and\ \citenamefont
  {Reddy}}]{Gandolfi:2011xu}%
  \BibitemOpen
  \bibfield  {author} {\bibinfo {author} {\bibfnamefont {S.}~\bibnamefont
  {Gandolfi}}, \bibinfo {author} {\bibfnamefont {J.}~\bibnamefont {Carlson}}, \
  and\ \bibinfo {author} {\bibfnamefont {S.}~\bibnamefont {Reddy}},\ }\href
  {\doibase 10.1103/PhysRevC.85.032801} {\bibfield  {journal} {\bibinfo
  {journal} {Phys. Rev. C}\ }\textbf {\bibinfo {volume} {85}},\ \bibinfo
  {pages} {032801} (\bibinfo {year} {2012})},\ \Eprint
  {http://arxiv.org/abs/1101.1921} {arXiv:1101.1921 [nucl-th]} \BibitemShut
  {NoStop}%
\bibitem [{\citenamefont {Masuda}\ \emph {et~al.}(2013)\citenamefont {Masuda},
  \citenamefont {Hatsuda},\ and\ \citenamefont {Takatsuka}}]{Masuda:2012ed}%
  \BibitemOpen
  \bibfield  {author} {\bibinfo {author} {\bibfnamefont {K.}~\bibnamefont
  {Masuda}}, \bibinfo {author} {\bibfnamefont {T.}~\bibnamefont {Hatsuda}}, \
  and\ \bibinfo {author} {\bibfnamefont {T.}~\bibnamefont {Takatsuka}},\ }\href
  {\doibase 10.1093/ptep/ptt045} {\bibfield  {journal} {\bibinfo  {journal}
  {PTEP}\ }\textbf {\bibinfo {volume} {2013}},\ \bibinfo {pages} {073D01}
  (\bibinfo {year} {2013})},\ \Eprint {http://arxiv.org/abs/1212.6803}
  {arXiv:1212.6803 [nucl-th]} \BibitemShut {NoStop}%
\bibitem [{\citenamefont {Kojo}\ \emph {et~al.}(2015)\citenamefont {Kojo},
  \citenamefont {Powell}, \citenamefont {Song},\ and\ \citenamefont
  {Baym}}]{Kojo:2014rca}%
  \BibitemOpen
  \bibfield  {author} {\bibinfo {author} {\bibfnamefont {T.}~\bibnamefont
  {Kojo}}, \bibinfo {author} {\bibfnamefont {P.~D.}\ \bibnamefont {Powell}},
  \bibinfo {author} {\bibfnamefont {Y.}~\bibnamefont {Song}}, \ and\ \bibinfo
  {author} {\bibfnamefont {G.}~\bibnamefont {Baym}},\ }\href {\doibase
  10.1103/PhysRevD.91.045003} {\bibfield  {journal} {\bibinfo  {journal} {Phys.
  Rev. D}\ }\textbf {\bibinfo {volume} {91}},\ \bibinfo {pages} {045003}
  (\bibinfo {year} {2015})},\ \Eprint {http://arxiv.org/abs/1412.1108}
  {arXiv:1412.1108 [hep-ph]} \BibitemShut {NoStop}%
\bibitem [{\citenamefont {Alford}\ \emph {et~al.}(2013)\citenamefont {Alford},
  \citenamefont {Han},\ and\ \citenamefont {Prakash}}]{Alford:2013aca}%
  \BibitemOpen
  \bibfield  {author} {\bibinfo {author} {\bibfnamefont {M.~G.}\ \bibnamefont
  {Alford}}, \bibinfo {author} {\bibfnamefont {S.}~\bibnamefont {Han}}, \ and\
  \bibinfo {author} {\bibfnamefont {M.}~\bibnamefont {Prakash}},\ }\href
  {\doibase 10.1103/PhysRevD.88.083013} {\bibfield  {journal} {\bibinfo
  {journal} {Phys. Rev. D}\ }\textbf {\bibinfo {volume} {88}},\ \bibinfo
  {pages} {083013} (\bibinfo {year} {2013})},\ \Eprint
  {http://arxiv.org/abs/1302.4732} {arXiv:1302.4732 [astro-ph.SR]} \BibitemShut
  {NoStop}%
\bibitem [{\citenamefont {Annala}\ \emph {et~al.}(2020)\citenamefont {Annala},
  \citenamefont {Gorda}, \citenamefont {Kurkela}, \citenamefont
  {N{\"a}ttil{\"a}},\ and\ \citenamefont {Vuorinen}}]{Annala:2019puf}%
  \BibitemOpen
  \bibfield  {author} {\bibinfo {author} {\bibfnamefont {E.}~\bibnamefont
  {Annala}}, \bibinfo {author} {\bibfnamefont {T.}~\bibnamefont {Gorda}},
  \bibinfo {author} {\bibfnamefont {A.}~\bibnamefont {Kurkela}}, \bibinfo
  {author} {\bibfnamefont {J.}~\bibnamefont {N{\"a}ttil{\"a}}}, \ and\ \bibinfo
  {author} {\bibfnamefont {A.}~\bibnamefont {Vuorinen}},\ }\href {\doibase
  10.1038/s41567-020-0914-9} {\bibfield  {journal} {\bibinfo  {journal} {Nature
  Phys.}\ }\textbf {\bibinfo {volume} {16}},\ \bibinfo {pages} {907} (\bibinfo
  {year} {2020})},\ \Eprint {http://arxiv.org/abs/1903.09121} {arXiv:1903.09121
  [astro-ph.HE]} \BibitemShut {NoStop}%
\bibitem [{\citenamefont {Annala}\ \emph {et~al.}(2023)\citenamefont {Annala},
  \citenamefont {Gorda}, \citenamefont {Hirvonen}, \citenamefont {Komoltsev},
  \citenamefont {Kurkela}, \citenamefont {N{\"a}ttil{\"a}},\ and\ \citenamefont
  {Vuorinen}}]{Annala:2023cwx}%
  \BibitemOpen
  \bibfield  {author} {\bibinfo {author} {\bibfnamefont {E.}~\bibnamefont
  {Annala}}, \bibinfo {author} {\bibfnamefont {T.}~\bibnamefont {Gorda}},
  \bibinfo {author} {\bibfnamefont {J.}~\bibnamefont {Hirvonen}}, \bibinfo
  {author} {\bibfnamefont {O.}~\bibnamefont {Komoltsev}}, \bibinfo {author}
  {\bibfnamefont {A.}~\bibnamefont {Kurkela}}, \bibinfo {author} {\bibfnamefont
  {J.}~\bibnamefont {N{\"a}ttil{\"a}}}, \ and\ \bibinfo {author} {\bibfnamefont
  {A.}~\bibnamefont {Vuorinen}},\ }\href {\doibase 10.1038/s41467-023-44051-y}
  {\bibfield  {journal} {\bibinfo  {journal} {Nature Commun.}\ }\textbf
  {\bibinfo {volume} {14}},\ \bibinfo {pages} {8451} (\bibinfo {year}
  {2023})},\ \Eprint {http://arxiv.org/abs/2303.11356} {arXiv:2303.11356
  [astro-ph.HE]} \BibitemShut {NoStop}%
\bibitem [{\citenamefont {Fujimoto}\ \emph {et~al.}(2022)\citenamefont
  {Fujimoto}, \citenamefont {Fukushima}, \citenamefont {McLerran},\ and\
  \citenamefont {Praszalowicz}}]{Fujimoto:2022ohj}%
  \BibitemOpen
  \bibfield  {author} {\bibinfo {author} {\bibfnamefont {Y.}~\bibnamefont
  {Fujimoto}}, \bibinfo {author} {\bibfnamefont {K.}~\bibnamefont {Fukushima}},
  \bibinfo {author} {\bibfnamefont {L.~D.}\ \bibnamefont {McLerran}}, \ and\
  \bibinfo {author} {\bibfnamefont {M.}~\bibnamefont {Praszalowicz}},\ }\href
  {\doibase 10.1103/PhysRevLett.129.252702} {\bibfield  {journal} {\bibinfo
  {journal} {Phys. Rev. Lett.}\ }\textbf {\bibinfo {volume} {129}},\ \bibinfo
  {pages} {252702} (\bibinfo {year} {2022})},\ \Eprint
  {http://arxiv.org/abs/2207.06753} {arXiv:2207.06753 [nucl-th]} \BibitemShut
  {NoStop}%
\bibitem [{\citenamefont {Takatsy}\ \emph {et~al.}(2023)\citenamefont
  {Takatsy}, \citenamefont {Kovacs}, \citenamefont {Wolf},\ and\ \citenamefont
  {Schaffner-Bielich}}]{Takatsy:2023xzf}%
  \BibitemOpen
  \bibfield  {author} {\bibinfo {author} {\bibfnamefont {J.}~\bibnamefont
  {Takatsy}}, \bibinfo {author} {\bibfnamefont {P.}~\bibnamefont {Kovacs}},
  \bibinfo {author} {\bibfnamefont {G.}~\bibnamefont {Wolf}}, \ and\ \bibinfo
  {author} {\bibfnamefont {J.}~\bibnamefont {Schaffner-Bielich}},\ }\href
  {\doibase 10.1103/PhysRevD.108.043002} {\bibfield  {journal} {\bibinfo
  {journal} {Phys. Rev. D}\ }\textbf {\bibinfo {volume} {108}},\ \bibinfo
  {pages} {043002} (\bibinfo {year} {2023})},\ \Eprint
  {http://arxiv.org/abs/2303.00013} {arXiv:2303.00013 [astro-ph.HE]}
  \BibitemShut {NoStop}%
\bibitem [{\citenamefont {Malik}\ \emph {et~al.}(2024)\citenamefont {Malik},
  \citenamefont {Dexheimer},\ and\ \citenamefont
  {Provid{\^e}ncia}}]{Malik:2024qjw}%
  \BibitemOpen
  \bibfield  {author} {\bibinfo {author} {\bibfnamefont {T.}~\bibnamefont
  {Malik}}, \bibinfo {author} {\bibfnamefont {V.}~\bibnamefont {Dexheimer}}, \
  and\ \bibinfo {author} {\bibfnamefont {C.}~\bibnamefont {Provid{\^e}ncia}},\
  }\href {\doibase 10.1103/PhysRevD.110.043042} {\bibfield  {journal} {\bibinfo
   {journal} {Phys. Rev. D}\ }\textbf {\bibinfo {volume} {110}},\ \bibinfo
  {pages} {043042} (\bibinfo {year} {2024})},\ \Eprint
  {http://arxiv.org/abs/2404.07936} {arXiv:2404.07936 [nucl-th]} \BibitemShut
  {NoStop}%
\bibitem [{\citenamefont {Panteleeva}\ \emph {et~al.}(2026)\citenamefont
  {Panteleeva}, \citenamefont {Alharazin},\ and\ \citenamefont
  {Epelbaum}}]{Panteleeva:2026zxs}%
  \BibitemOpen
  \bibfield  {author} {\bibinfo {author} {\bibfnamefont {J.~Y.}\ \bibnamefont
  {Panteleeva}}, \bibinfo {author} {\bibfnamefont {H.}~\bibnamefont
  {Alharazin}}, \ and\ \bibinfo {author} {\bibfnamefont {E.}~\bibnamefont
  {Epelbaum}},\ }\href@noop {} {\  (\bibinfo {year} {2026})},\ \Eprint
  {http://arxiv.org/abs/2608.17457} {arXiv:2608.17457 [nucl-th]} \BibitemShut
  {NoStop}%
\bibitem [{\citenamefont {Baym}\ \emph {et~al.}(1971)\citenamefont {Baym},
  \citenamefont {Pethick},\ and\ \citenamefont {Sutherland}}]{Baym:1971pw}%
  \BibitemOpen
  \bibfield  {author} {\bibinfo {author} {\bibfnamefont {G.}~\bibnamefont
  {Baym}}, \bibinfo {author} {\bibfnamefont {C.}~\bibnamefont {Pethick}}, \
  and\ \bibinfo {author} {\bibfnamefont {P.}~\bibnamefont {Sutherland}},\
  }\href {\doibase 10.1086/151216} {\bibfield  {journal} {\bibinfo  {journal}
  {Astrophys. J.}\ }\textbf {\bibinfo {volume} {170}},\ \bibinfo {pages} {299}
  (\bibinfo {year} {1971})}\BibitemShut {NoStop}%
\bibitem [{\citenamefont {Ma}\ \emph {et~al.}(2026)\citenamefont {Ma},
  \citenamefont {Ma},\ and\ \citenamefont {Xiong}}]{Ma:2026kun}%
  \BibitemOpen
  \bibfield  {author} {\bibinfo {author} {\bibfnamefont {Y.}~\bibnamefont
  {Ma}}, \bibinfo {author} {\bibfnamefont {Y.-L.}\ \bibnamefont {Ma}}, \ and\
  \bibinfo {author} {\bibfnamefont {J.-Y.}\ \bibnamefont {Xiong}},\ }\href@noop
  {} {\  (\bibinfo {year} {2026})},\ \Eprint {http://arxiv.org/abs/2603.01933}
  {arXiv:2603.01933 [nucl-th]} \BibitemShut {NoStop}%
\bibitem [{\citenamefont {Piekarewicz}(2006)}]{Piekarewicz:2006bw}%
  \BibitemOpen
  \bibfield  {author} {\bibinfo {author} {\bibfnamefont {J.}~\bibnamefont
  {Piekarewicz}},\ }in\ \href@noop {} {\emph {\bibinfo {booktitle}
  {{International Conference on Current Problems of Nuclear Physics and Atomic
  Energy (NPAE 2006)}}}}\ (\bibinfo {year} {2006})\ pp.\ \bibinfo {pages}
  {33--42},\ \Eprint {http://arxiv.org/abs/nucl-th/0607039}
  {arXiv:nucl-th/0607039} \BibitemShut {NoStop}%
\bibitem [{\citenamefont {Sugahara}\ and\ \citenamefont
  {Toki}(1994)}]{Sugahara:1993wz}%
  \BibitemOpen
  \bibfield  {author} {\bibinfo {author} {\bibfnamefont {Y.}~\bibnamefont
  {Sugahara}}\ and\ \bibinfo {author} {\bibfnamefont {H.}~\bibnamefont
  {Toki}},\ }\href {\doibase 10.1016/0375-9474(94)90923-7} {\bibfield
  {journal} {\bibinfo  {journal} {Nucl. Phys. A}\ }\textbf {\bibinfo {volume}
  {579}},\ \bibinfo {pages} {557} (\bibinfo {year} {1994})}\BibitemShut
  {NoStop}%
\bibitem [{\citenamefont {Lalazissis}\ \emph {et~al.}(1997)\citenamefont
  {Lalazissis}, \citenamefont {Konig},\ and\ \citenamefont
  {Ring}}]{Lalazissis:1996rd}%
  \BibitemOpen
  \bibfield  {author} {\bibinfo {author} {\bibfnamefont {G.~A.}\ \bibnamefont
  {Lalazissis}}, \bibinfo {author} {\bibfnamefont {J.}~\bibnamefont {Konig}}, \
  and\ \bibinfo {author} {\bibfnamefont {P.}~\bibnamefont {Ring}},\ }\href
  {\doibase 10.1103/PhysRevC.55.540} {\bibfield  {journal} {\bibinfo  {journal}
  {Phys. Rev. C}\ }\textbf {\bibinfo {volume} {55}},\ \bibinfo {pages} {540}
  (\bibinfo {year} {1997})},\ \Eprint {http://arxiv.org/abs/nucl-th/9607039}
  {arXiv:nucl-th/9607039} \BibitemShut {NoStop}%
\bibitem [{\citenamefont {Li}\ \emph {et~al.}(2022)\citenamefont {Li},
  \citenamefont {Cai}, \citenamefont {Zhou}, \citenamefont {Jiang},\ and\
  \citenamefont {Chen}}]{Li:2022okx}%
  \BibitemOpen
  \bibfield  {author} {\bibinfo {author} {\bibfnamefont {F.}~\bibnamefont
  {Li}}, \bibinfo {author} {\bibfnamefont {B.-J.}\ \bibnamefont {Cai}},
  \bibinfo {author} {\bibfnamefont {Y.}~\bibnamefont {Zhou}}, \bibinfo {author}
  {\bibfnamefont {W.-Z.}\ \bibnamefont {Jiang}}, \ and\ \bibinfo {author}
  {\bibfnamefont {L.-W.}\ \bibnamefont {Chen}},\ }\href {\doibase
  10.3847/1538-4357/ac5e2a} {\bibfield  {journal} {\bibinfo  {journal}
  {Astrophys. J.}\ }\textbf {\bibinfo {volume} {929}},\ \bibinfo {pages} {183}
  (\bibinfo {year} {2022})},\ \Eprint {http://arxiv.org/abs/2202.08705}
  {arXiv:2202.08705 [nucl-th]} \BibitemShut {NoStop}%
\bibitem [{\citenamefont {Zhang}\ \emph {et~al.}(2026)\citenamefont {Zhang},
  \citenamefont {Ma},\ and\ \citenamefont {Ma}}]{Zhang:2024iye}%
  \BibitemOpen
  \bibfield  {author} {\bibinfo {author} {\bibfnamefont {L.-Q.}\ \bibnamefont
  {Zhang}}, \bibinfo {author} {\bibfnamefont {Y.}~\bibnamefont {Ma}}, \ and\
  \bibinfo {author} {\bibfnamefont {Y.-L.}\ \bibnamefont {Ma}},\ }\href
  {\doibase 10.1088/1674-1137/ae18ae} {\bibfield  {journal} {\bibinfo
  {journal} {Chin. Phys. C}\ }\textbf {\bibinfo {volume} {50}},\ \bibinfo
  {pages} {024108} (\bibinfo {year} {2026})},\ \Eprint
  {http://arxiv.org/abs/2412.19023} {arXiv:2412.19023 [nucl-th]} \BibitemShut
  {NoStop}%
\bibitem [{\citenamefont {Alford}\ \emph {et~al.}(2022)\citenamefont {Alford},
  \citenamefont {Brodie}, \citenamefont {Haber},\ and\ \citenamefont
  {Tews}}]{Alford:2022bpp}%
  \BibitemOpen
  \bibfield  {author} {\bibinfo {author} {\bibfnamefont {M.~G.}\ \bibnamefont
  {Alford}}, \bibinfo {author} {\bibfnamefont {L.}~\bibnamefont {Brodie}},
  \bibinfo {author} {\bibfnamefont {A.}~\bibnamefont {Haber}}, \ and\ \bibinfo
  {author} {\bibfnamefont {I.}~\bibnamefont {Tews}},\ }\href {\doibase
  10.1103/PhysRevC.106.055804} {\bibfield  {journal} {\bibinfo  {journal}
  {Phys. Rev. C}\ }\textbf {\bibinfo {volume} {106}},\ \bibinfo {pages}
  {055804} (\bibinfo {year} {2022})},\ \Eprint
  {http://arxiv.org/abs/2205.10283} {arXiv:2205.10283 [nucl-th]} \BibitemShut
  {NoStop}%
\bibitem [{\citenamefont {G{\"o}ttling}\ \emph {et~al.}(2026)\citenamefont
  {G{\"o}ttling}, \citenamefont {Hoff}, \citenamefont {Hebeler},\ and\
  \citenamefont {Schwenk}}]{Gottling:2025ohe}%
  \BibitemOpen
  \bibfield  {author} {\bibinfo {author} {\bibfnamefont {H.}~\bibnamefont
  {G{\"o}ttling}}, \bibinfo {author} {\bibfnamefont {L.}~\bibnamefont {Hoff}},
  \bibinfo {author} {\bibfnamefont {K.}~\bibnamefont {Hebeler}}, \ and\
  \bibinfo {author} {\bibfnamefont {A.}~\bibnamefont {Schwenk}},\ }\href
  {\doibase 10.1103/t9y8-nbf4} {\bibfield  {journal} {\bibinfo  {journal}
  {Phys. Rev. C}\ }\textbf {\bibinfo {volume} {113}},\ \bibinfo {pages}
  {055804} (\bibinfo {year} {2026})},\ \Eprint
  {http://arxiv.org/abs/2512.19593} {arXiv:2512.19593 [nucl-th]} \BibitemShut
  {NoStop}%
\bibitem [{\citenamefont {Lattimer}\ and\ \citenamefont
  {Prakash}(2001)}]{Lattimer:2000nx}%
  \BibitemOpen
  \bibfield  {author} {\bibinfo {author} {\bibfnamefont {J.~M.}\ \bibnamefont
  {Lattimer}}\ and\ \bibinfo {author} {\bibfnamefont {M.}~\bibnamefont
  {Prakash}},\ }\href {\doibase 10.1086/319702} {\bibfield  {journal} {\bibinfo
   {journal} {Astrophys. J.}\ }\textbf {\bibinfo {volume} {550}},\ \bibinfo
  {pages} {426} (\bibinfo {year} {2001})},\ \Eprint
  {http://arxiv.org/abs/astro-ph/0002232} {arXiv:astro-ph/0002232} \BibitemShut
  {NoStop}%
\bibitem [{\citenamefont {Rehberg}\ \emph {et~al.}(1996)\citenamefont
  {Rehberg}, \citenamefont {Klevansky},\ and\ \citenamefont
  {Hufner}}]{Rehberg:1995kh}%
  \BibitemOpen
  \bibfield  {author} {\bibinfo {author} {\bibfnamefont {P.}~\bibnamefont
  {Rehberg}}, \bibinfo {author} {\bibfnamefont {S.~P.}\ \bibnamefont
  {Klevansky}}, \ and\ \bibinfo {author} {\bibfnamefont {J.}~\bibnamefont
  {Hufner}},\ }\href {\doibase 10.1103/PhysRevC.53.410} {\bibfield  {journal}
  {\bibinfo  {journal} {Phys. Rev. C}\ }\textbf {\bibinfo {volume} {53}},\
  \bibinfo {pages} {410} (\bibinfo {year} {1996})},\ \Eprint
  {http://arxiv.org/abs/hep-ph/9506436} {arXiv:hep-ph/9506436} \BibitemShut
  {NoStop}%
\bibitem [{\citenamefont {McLerran}\ and\ \citenamefont
  {Reddy}(2019)}]{McLerran:2018hbz}%
  \BibitemOpen
  \bibfield  {author} {\bibinfo {author} {\bibfnamefont {L.}~\bibnamefont
  {McLerran}}\ and\ \bibinfo {author} {\bibfnamefont {S.}~\bibnamefont
  {Reddy}},\ }\href {\doibase 10.1103/PhysRevLett.122.122701} {\bibfield
  {journal} {\bibinfo  {journal} {Phys. Rev. Lett.}\ }\textbf {\bibinfo
  {volume} {122}},\ \bibinfo {pages} {122701} (\bibinfo {year} {2019})},\
  \Eprint {http://arxiv.org/abs/1811.12503} {arXiv:1811.12503 [nucl-th]}
  \BibitemShut {NoStop}%
\bibitem [{\citenamefont {Kini}\ \emph {et~al.}(2026)\citenamefont {Kini} \emph
  {et~al.}}]{Kini:2026rjx}%
  \BibitemOpen
  \bibfield  {author} {\bibinfo {author} {\bibfnamefont {Y.}~\bibnamefont
  {Kini}} \emph {et~al.},\ }\href {\doibase 10.3847/1538-4357/ae733e}
  {\bibfield  {journal} {\bibinfo  {journal} {Astrophys. J.}\ }\textbf
  {\bibinfo {volume} {1005}},\ \bibinfo {pages} {201} (\bibinfo {year}
  {2026})},\ \Eprint {http://arxiv.org/abs/2602.23743} {arXiv:2602.23743
  [astro-ph.HE]} \BibitemShut {NoStop}%
\bibitem [{\citenamefont {Salmi}\ \emph {et~al.}(2024)\citenamefont {Salmi}
  \emph {et~al.}}]{Salmi:2024aum}%
  \BibitemOpen
  \bibfield  {author} {\bibinfo {author} {\bibfnamefont {T.}~\bibnamefont
  {Salmi}} \emph {et~al.},\ }\href {\doibase 10.3847/1538-4357/ad5f1f}
  {\bibfield  {journal} {\bibinfo  {journal} {Astrophys. J.}\ }\textbf
  {\bibinfo {volume} {974}},\ \bibinfo {pages} {294} (\bibinfo {year}
  {2024})},\ \Eprint {http://arxiv.org/abs/2406.14466} {arXiv:2406.14466
  [astro-ph.HE]} \BibitemShut {NoStop}%
\bibitem [{\citenamefont {Arzoumanian}\ \emph
  {et~al.}(2018{\natexlab{a}})\citenamefont {Arzoumanian} \emph
  {et~al.}}]{NANOGRAV:2018hou}%
  \BibitemOpen
  \bibfield  {author} {\bibinfo {author} {\bibfnamefont {Z.}~\bibnamefont
  {Arzoumanian}} \emph {et~al.} (\bibinfo {collaboration} {NANOGRAV}),\ }\href
  {\doibase 10.3847/1538-4357/aabd3b} {\bibfield  {journal} {\bibinfo
  {journal} {Astrophys. J.}\ }\textbf {\bibinfo {volume} {859}},\ \bibinfo
  {pages} {47} (\bibinfo {year} {2018}{\natexlab{a}})},\ \Eprint
  {http://arxiv.org/abs/1801.02617} {arXiv:1801.02617 [astro-ph.HE]}
  \BibitemShut {NoStop}%
\bibitem [{\citenamefont {Todd-Rutel}\ and\ \citenamefont
  {Piekarewicz}(2005)}]{Todd-Rutel:2005yzo}%
  \BibitemOpen
  \bibfield  {author} {\bibinfo {author} {\bibfnamefont {B.~G.}\ \bibnamefont
  {Todd-Rutel}}\ and\ \bibinfo {author} {\bibfnamefont {J.}~\bibnamefont
  {Piekarewicz}},\ }\href {\doibase 10.1103/PhysRevLett.95.122501} {\bibfield
  {journal} {\bibinfo  {journal} {Phys. Rev. Lett.}\ }\textbf {\bibinfo
  {volume} {95}},\ \bibinfo {pages} {122501} (\bibinfo {year} {2005})},\
  \Eprint {http://arxiv.org/abs/nucl-th/0504034} {arXiv:nucl-th/0504034}
  \BibitemShut {NoStop}%
\bibitem [{\citenamefont {Hatsuda}\ and\ \citenamefont
  {Kunihiro}(1994)}]{Hatsuda:1994pi}%
  \BibitemOpen
  \bibfield  {author} {\bibinfo {author} {\bibfnamefont {T.}~\bibnamefont
  {Hatsuda}}\ and\ \bibinfo {author} {\bibfnamefont {T.}~\bibnamefont
  {Kunihiro}},\ }\href {\doibase 10.1016/0370-1573(94)90022-1} {\bibfield
  {journal} {\bibinfo  {journal} {Phys. Rept.}\ }\textbf {\bibinfo {volume}
  {247}},\ \bibinfo {pages} {221} (\bibinfo {year} {1994})},\ \Eprint
  {http://arxiv.org/abs/hep-ph/9401310} {arXiv:hep-ph/9401310} \BibitemShut
  {NoStop}%
\bibitem [{\citenamefont {Ratti}\ \emph {et~al.}(2006)\citenamefont {Ratti},
  \citenamefont {Thaler},\ and\ \citenamefont {Weise}}]{Ratti:2006ka}%
  \BibitemOpen
  \bibfield  {author} {\bibinfo {author} {\bibfnamefont {C.}~\bibnamefont
  {Ratti}}, \bibinfo {author} {\bibfnamefont {M.~A.}\ \bibnamefont {Thaler}}, \
  and\ \bibinfo {author} {\bibfnamefont {W.}~\bibnamefont {Weise}},\
  }\href@noop {} {\bibfield  {journal} {\bibinfo  {journal} {Rom. Rep. Phys.}\
  }\textbf {\bibinfo {volume} {58}},\ \bibinfo {pages} {13} (\bibinfo {year}
  {2006})}\BibitemShut {NoStop}%
\bibitem [{\citenamefont {Arzoumanian}\ \emph
  {et~al.}(2018{\natexlab{b}})\citenamefont {Arzoumanian} \emph
  {et~al.}}]{NANOGrav:2017wvv}%
  \BibitemOpen
  \bibfield  {author} {\bibinfo {author} {\bibfnamefont {Z.}~\bibnamefont
  {Arzoumanian}} \emph {et~al.} (\bibinfo {collaboration} {NANOGrav}),\ }\href
  {\doibase 10.3847/1538-4365/aab5b0} {\bibfield  {journal} {\bibinfo
  {journal} {Astrophys. J. Suppl.}\ }\textbf {\bibinfo {volume} {235}},\
  \bibinfo {pages} {37} (\bibinfo {year} {2018}{\natexlab{b}})},\ \Eprint
  {http://arxiv.org/abs/1801.01837} {arXiv:1801.01837 [astro-ph.HE]}
  \BibitemShut {NoStop}%
\bibitem [{\citenamefont {Dutra}\ \emph {et~al.}(2014)\citenamefont {Dutra},
  \citenamefont {Louren{\c{c}}o}, \citenamefont {Avancini}, \citenamefont
  {Carlson}, \citenamefont {Delfino}, \citenamefont {Menezes}, \citenamefont
  {Provid{\^e}ncia}, \citenamefont {Typel},\ and\ \citenamefont
  {Stone}}]{Dutra:2014qga}%
  \BibitemOpen
  \bibfield  {author} {\bibinfo {author} {\bibfnamefont {M.}~\bibnamefont
  {Dutra}}, \bibinfo {author} {\bibfnamefont {O.}~\bibnamefont
  {Louren{\c{c}}o}}, \bibinfo {author} {\bibfnamefont {S.~S.}\ \bibnamefont
  {Avancini}}, \bibinfo {author} {\bibfnamefont {B.~V.}\ \bibnamefont
  {Carlson}}, \bibinfo {author} {\bibfnamefont {A.}~\bibnamefont {Delfino}},
  \bibinfo {author} {\bibfnamefont {D.~P.}\ \bibnamefont {Menezes}}, \bibinfo
  {author} {\bibfnamefont {C.}~\bibnamefont {Provid{\^e}ncia}}, \bibinfo
  {author} {\bibfnamefont {S.}~\bibnamefont {Typel}}, \ and\ \bibinfo {author}
  {\bibfnamefont {J.~R.}\ \bibnamefont {Stone}},\ }\href {\doibase
  10.1103/PhysRevC.90.055203} {\bibfield  {journal} {\bibinfo  {journal} {Phys.
  Rev. C}\ }\textbf {\bibinfo {volume} {90}},\ \bibinfo {pages} {055203}
  (\bibinfo {year} {2014})},\ \Eprint {http://arxiv.org/abs/1405.3633}
  {arXiv:1405.3633 [nucl-th]} \BibitemShut {NoStop}%
\bibitem [{\citenamefont {Lattimer}\ and\ \citenamefont
  {Lim}(2013)}]{Lattimer:2012xj}%
  \BibitemOpen
  \bibfield  {author} {\bibinfo {author} {\bibfnamefont {J.~M.}\ \bibnamefont
  {Lattimer}}\ and\ \bibinfo {author} {\bibfnamefont {Y.}~\bibnamefont {Lim}},\
  }\href {\doibase 10.1088/0004-637X/771/1/51} {\bibfield  {journal} {\bibinfo
  {journal} {Astrophys. J.}\ }\textbf {\bibinfo {volume} {771}},\ \bibinfo
  {pages} {51} (\bibinfo {year} {2013})},\ \Eprint
  {http://arxiv.org/abs/1203.4286} {arXiv:1203.4286 [nucl-th]} \BibitemShut
  {NoStop}%
\bibitem [{\citenamefont {Danielewicz}\ and\ \citenamefont
  {Lee}(2014)}]{Danielewicz:2013upa}%
  \BibitemOpen
  \bibfield  {author} {\bibinfo {author} {\bibfnamefont {P.}~\bibnamefont
  {Danielewicz}}\ and\ \bibinfo {author} {\bibfnamefont {J.}~\bibnamefont
  {Lee}},\ }\href {\doibase 10.1016/j.nuclphysa.2013.11.005} {\bibfield
  {journal} {\bibinfo  {journal} {Nucl. Phys. A}\ }\textbf {\bibinfo {volume}
  {922}},\ \bibinfo {pages} {1} (\bibinfo {year} {2014})},\ \Eprint
  {http://arxiv.org/abs/1307.4130} {arXiv:1307.4130 [nucl-th]} \BibitemShut
  {NoStop}%
\bibitem [{\citenamefont {Dutra}\ \emph {et~al.}(2012)\citenamefont {Dutra},
  \citenamefont {Lourenco}, \citenamefont {Sa~Martins}, \citenamefont
  {Delfino}, \citenamefont {Stone},\ and\ \citenamefont
  {Stevenson}}]{Dutra:2012mb}%
  \BibitemOpen
  \bibfield  {author} {\bibinfo {author} {\bibfnamefont {M.}~\bibnamefont
  {Dutra}}, \bibinfo {author} {\bibfnamefont {O.}~\bibnamefont {Lourenco}},
  \bibinfo {author} {\bibfnamefont {J.~S.}\ \bibnamefont {Sa~Martins}},
  \bibinfo {author} {\bibfnamefont {A.}~\bibnamefont {Delfino}}, \bibinfo
  {author} {\bibfnamefont {J.~R.}\ \bibnamefont {Stone}}, \ and\ \bibinfo
  {author} {\bibfnamefont {P.~D.}\ \bibnamefont {Stevenson}},\ }\href {\doibase
  10.1103/PhysRevC.85.035201} {\bibfield  {journal} {\bibinfo  {journal} {Phys.
  Rev. C}\ }\textbf {\bibinfo {volume} {85}},\ \bibinfo {pages} {035201}
  (\bibinfo {year} {2012})},\ \Eprint {http://arxiv.org/abs/1202.3902}
  {arXiv:1202.3902 [nucl-th]} \BibitemShut {NoStop}%
\bibitem [{\citenamefont {Cai}\ and\ \citenamefont {Chen}(2017)}]{Cai:2014kya}%
  \BibitemOpen
  \bibfield  {author} {\bibinfo {author} {\bibfnamefont {B.-J.}\ \bibnamefont
  {Cai}}\ and\ \bibinfo {author} {\bibfnamefont {L.-W.}\ \bibnamefont {Chen}},\
  }\href {\doibase 10.1007/s41365-017-0329-1} {\bibfield  {journal} {\bibinfo
  {journal} {Nucl. Sci. Tech.}\ }\textbf {\bibinfo {volume} {28}},\ \bibinfo
  {pages} {185} (\bibinfo {year} {2017})},\ \Eprint
  {http://arxiv.org/abs/1402.4242} {arXiv:1402.4242 [nucl-th]} \BibitemShut
  {NoStop}%
\bibitem [{\citenamefont {Xie}\ and\ \citenamefont {Li}(2021)}]{Xie:2020kta}%
  \BibitemOpen
  \bibfield  {author} {\bibinfo {author} {\bibfnamefont {W.-J.}\ \bibnamefont
  {Xie}}\ and\ \bibinfo {author} {\bibfnamefont {B.-A.}\ \bibnamefont {Li}},\
  }\href {\doibase 10.1088/1361-6471/abd25a} {\bibfield  {journal} {\bibinfo
  {journal} {J. Phys. G}\ }\textbf {\bibinfo {volume} {48}},\ \bibinfo {pages}
  {025110} (\bibinfo {year} {2021})},\ \Eprint
  {http://arxiv.org/abs/2001.03669} {arXiv:2001.03669 [nucl-th]} \BibitemShut
  {NoStop}%
\bibitem [{\citenamefont {Margueron}\ \emph {et~al.}(2018)\citenamefont
  {Margueron}, \citenamefont {Hoffmann~Casali},\ and\ \citenamefont
  {Gulminelli}}]{Margueron:2017eqc}%
  \BibitemOpen
  \bibfield  {author} {\bibinfo {author} {\bibfnamefont {J.}~\bibnamefont
  {Margueron}}, \bibinfo {author} {\bibfnamefont {R.}~\bibnamefont
  {Hoffmann~Casali}}, \ and\ \bibinfo {author} {\bibfnamefont {F.}~\bibnamefont
  {Gulminelli}},\ }\href {\doibase 10.1103/PhysRevC.97.025805} {\bibfield
  {journal} {\bibinfo  {journal} {Phys. Rev. C}\ }\textbf {\bibinfo {volume}
  {97}},\ \bibinfo {pages} {025805} (\bibinfo {year} {2018})},\ \Eprint
  {http://arxiv.org/abs/1708.06894} {arXiv:1708.06894 [nucl-th]} \BibitemShut
  {NoStop}%
\end{thebibliography}%

\end{document}